\documentclass[twocolumn,showpacs,amsmath,amssymb,superscriptaddress]{revtex4-1}
\newcommand{\fant}[1]{\phantom{#1}}
\newcommand{\be}{\begin{equation}}
\newcommand{\ee}{\end{equation}}
\newcommand{\wdg}{\wedge}
\newcommand{\ot}{\otimes}
\usepackage{slashed}
\usepackage{amsmath}
\usepackage{amsfonts}
\usepackage{amssymb}

\begin{document}

\begin{abstract}

A general  scheme of constructing scalar-tensor equivalents to modified gravitational actions are studied using the algebra of exterior differential forms and the first order formalism that allows an independent connection and coframe. By introducing appropriate constraints on the connection, pseudo-Riemannian cases as well as non-Riemannian cases are discussed for various gravitational models. The issue of the dynamical degree  of freedom for the resulting scalar fields is discussed at the level of the field equations. Explicit scalar-tensor equivalents for gravitational models  based on $f(R)$ models, the quadratic curvature Lagrangians  and the models involving the gradients of the scalar curvature are presented. In particular, explicit scalar-tensor equivalence for gravitational Lagrangians  popular in some cosmological models are constructed.

\end{abstract}

\title{Multi-scalar-tensor equivalents for modified gravitational actions}

\author{Ahmet Baykal}
\email{abaykal@nigde.edu.tr}
\affiliation{Department of Physics, Faculty of Science and Letters, Ni\u gde University,  51240 Ni\u gde, Turkey}
\author{\"Ozg\"ur Delice}
\email{ozgur.delice@marmara.edu.tr}
\affiliation{Department of Physics, Faculty of Science and Letters, Marmara University, 34722 \.Istanbul, Turkey}
\date\today

\maketitle

\date{\today}

\section{Introduction}
 Einstein's theory of general relativity is a well tested theory explaining the gravitational interaction ranging from weak to strong fields and  from solar system scale to the whole Universe. However, its alternatives are taking a lot of interest recently. For an extensive survey of several motivations and various aspects of modified gravity theories, we refer to the  recent reviews \cite{clifton-phys-rep,faraoni} and the references therein.
 Actually, apart from mathematical curiosity to understand mathematical and physical properties of these theories by considering and comparing to the alternative theories, there are several physical motivations. For example,
the quantum effects of gravity  seem to require higher order curvature correction terms in the theory. In particular, the gravitational models based on scalars constructed from the terms that are quadratic in curvature components 
are motivated in different contexts ranging from low energy
 limit of string theories \cite{zweibach}, quantum theory of gravity \cite{starobinski,stelle}, to viable cosmological models \cite{hj-scmidt-history}.
 One of the popular alternative theories is based on  $f(R)$ Lagrangian which replaces the usual Ricci scalar term $R$ in the Einstein-Hilbert action with an arbitrary algebraic function of $R$.   The theories with  corresponding Lagrangian involving functions of  other curvature scalars or collections of those scalars,  in Riemannian or more general contexts are also studied extensively.

An important
feature of modified gravitational Lagrangians is the concept of scalar-tensor (ST) equivalence for these models. For instance, $f(R)$ theory is known to have equivalent  Brans-Dicke-type scalar-tensor theory with an extra potential term for the scalar field.  Actually, scalar-tensor theories 
 that are equivalent to particular $f(R)$ models were first introduced long ago \cite{whitt,teyssandier} and since then it is reintroduced, studied and has been made use of intermittently to this day.
ST equivalent of $f(R)$ models in different approaches has been introduced both in metric and in Palatini approaches. In particular, in a first order theory where connection and metric are treated as independent,   in the Brans-Dicke-type theories \cite{brans-dicke}, the scalar field is known to generate an algebraic torsion \cite{dereli-tucker} and thus it can be cast into pure metric theory by eliminating torsion \cite{sotiriou-torsion}. By means of the ST equivalence,  these considerations  also apply to generic $f(R)$ models as well.  The ST equivalence of $f(R)$ models has also
helped to investigate various aspects of $f(R)$ theories, such as the chameleon mechanism in  $f(R)$ theories \cite{chameleon}, 
the study  of  a Birkoff-Jebsen  like theorem in generic $f(R)$ theories \cite{capozziello-birkhoff,faraoni-birkhoff}, and
the search for  gravitational-wave solutions of modified gravity models \cite{corda}.

In higher  order metric theories of gravity,  the Legendre transformation was previously introduced  in \cite{magnano-grg, magnano-sokolowski} for Lagrangian densities depending on scalar curvature and in particular Ricci tensor in a nonlinear way. Later, this work is extended to study the general quadratic
curvature Lagrangian density of the form
\be
\mathcal{L}
=
(aR^{2}+bR^{\alpha\beta}R_{\alpha\beta}+cR^{\alpha\beta\mu\nu}R_{\alpha\beta\mu\nu})*1,
\ee
where the Legendre transformation with respect to tensorial quantities was introduced in a more general mathematical setting \cite{magnano-cqg}.
Later, the ST equivalence is extended to Palatini-type modified gravity theories
 where metric and connection are treated as independent gravitational variables \cite{flanagan}. More recently, the Legendre transformation of modified gravity is revisited \cite{saltas} in the context of ST equivalents of  models based on $R+f(G)$ \cite{NojiriOdintsovFG}, to study  Gibbons-Hawking surface terms of dynamically equivalent theories.

Meanwhile, the idea of multi-scalar-tensor equivalents were reintroduced in the study of  particle content of particular modified gravitational models based on $f(R, R_{\mu\nu}R^{\mu\nu}, R_{\mu\nu\alpha\beta}R^{\mu\nu\alpha\beta})$, to show that such modified theories share the common undesirable feature of spin-2 ghosts with generic quadratic curvature models \cite{chiba}.  The present work can be considered as an extension  of the multi-scalar-tensor equivalents, which are presented at the level of action, to the level of field equations for the physically relevant case studied in
\cite{chiba} as well as others.  Thus,  one of the motivations for the present work is to show that
the mathematical formulation of the Brans-Dicke type ST equivalence for generic  $f(R)$ theories  can naturally be extended to obtain the ST equivalents for a wider set of modified Lagrangians based on arbitrary curvature scalars in the general form indicated above. Such extensions are presented explicitly using, in particular, the curvature scalars built out of the quadratic curvature scalars for the sake of the simplicity of the presentation as well as regarding the application to general theory of relativity.

The multi-scalar-tensor equivalence for generic modified Lagrangians is studied extensively in \cite{starobinski-shapiro} regarding the number of scalar fields involved where the multi-scalar-tensor equivalents are constructed  using the properties of the Legendre transform with constraints.
As will explicitly be illustrated in the study of ST equivalents for various gravitational actions below, the dynamical degree of  freedom in  an ST equivalent for a gravitational model based on a function of a collection of several curvature scalars  depends on the  form of the function
and the total number of independent scalar fields is related to the rank of the Hessian matrix of the Legendre transformation used in the construction of ST equivalents \cite{starobinski-shapiro}.
The present work, on the other hand,  addresses the issue of the dynamical degree of freedom corresponding to the scalar fields in the resulting ST-equivalent model. For a given modified gravitational Lagrangian, the dynamical degree of freedom carried by the scalar fields can be studied at the level of the corresponding
field equations and the dynamics of the scalar fields is not apparent at the level of action. For this reason,
some  explicit models involving modified quadratic curvature Lagrangians are presented since such
gravitational models have important cosmological features.  For example, they provide a framework for inflation \cite{starobinski}, or else, in models with nonlinear curvature terms, it is possible to avoid cosmological singularity \cite{kerner}. Later, versions of these models are also studied  in order to account for the
late-time acceleration of the Universe as a viable alternative to dark energy models \cite{carroll2}.

The outline of the paper is as follows.
In the next section,  the first order constrained formalism  is reviewed using the algebra of the exterior forms relative to an orthonormal coframe. The exterior differential form notation and the geometrical conventions used  are adopted  from \cite{straumann,tucker,hehl}.  In the third section, some important features of the scalar-tensor equivalence of $f(R)$ theory are briefly presented in both pseudo-Riemannian and non-Riemannian context with an independent connection having a nonvanishing torsion and a nonmetricity.
The ST equivalence is then slightly generalized to a  generic modified Lagrangian depending on an arbitrary contraction of Riemann tensor
and the equivalence is presented at the level of the field equations and also a criterion for the resulting scalar field to be dynamical is given following the  study of the Lagrange multiplier term of the generic modified Lagrangian. In the fourth section, the ST equivalence of gravitational Lagrangians of the form $f(R,  Q, P, K, S)$ is presented in some generality with $Q, P, K, S$ being the quadratic curvature scalars defined below. Subsequently, ST equivalence of various modified gravitational Lagrangians, such as modified Ricci or Weyl square gravity, which could have important applications to cosmology and to the study of black holes are investigated in detail as applications of this equivalence scheme.
Moreover, modified gravitational Lagrangians popular in cosmology involving inverse powers of quadratic curvature terms are
shown to be dynamically equivalent to simpler quadratic curvature gravity models with nonminimal scalar  couplings. As the last application,
the ST equivalence for modified  6th order  gravitational  Lagrangians involving the derivatives of the scalar curvature is presented.
The paper concludes with  brief comments on the general features of the ST equivalence. A discussion of the relation between the equations we have discussed in the orthonormal coframe and their coordinate frame expressions is presented in the Appendix.

\section{Field Equations relative  to an Orthonormal Coframe}

In this section the  scheme  of calculation of variational derivatives relative to an orthonormal coframe will briefly be presented
 \cite{kopczynsky,baykal-delice,baykal,var-dereli-tucker}. In the subsequent  sections,  the general formulas of this section will be applied to all the modified gravitational  actions considered below.
Using the language of differential forms defined on pseudo-Riemannian manifolds, the basic independent gravitational variables that will be used below are defined in this section. In terms of the local basis, coframe 1-forms are denoted by $\{\theta^\alpha\}$ and the metric  tensor takes the form
$g=\eta_{\alpha\beta}\theta^\alpha\ot\theta^\beta$ with $\eta_{\alpha\beta}=\mbox{diag}(-+++)$. The exterior product of the basis 1-forms will be  abbreviated as
$\theta^{\alpha}\wdg \theta^\beta\wdg\cdots\equiv\theta^{\alpha\beta\cdots}$. The covariant exterior derivative $D$ acts on tensor-valued forms.
The torsion 2-form $\Theta^\alpha=\frac{1}{2}T^{\alpha}_{\fant{a}\beta\mu}\theta^{\beta\mu}$ can be  defined as
\be\label{cartan-structure-eqn1}
\Theta^\alpha=D\theta^\alpha=d\theta^\alpha+\omega^{\alpha}_{\fant{a}\beta}\wdg\theta^\beta,
\ee
where $T^{\alpha}_{\fant{a}\beta\mu}$ are the components of the torsion tensor and $\omega^{\alpha}_{\fant{a}\beta}$ are the connection 1-forms. Relative to an orthonormal coframe, metric compatibility for the connection 1-forms reads $\omega_{\alpha\beta}+\omega_{\beta\alpha}=0$. In terms of the curvature 2-forms $\Omega^{\alpha}_{\fant{a}\beta}$, the Cartan's second structure equation has the form
\be
\Omega^{\alpha}_{\fant{a}\beta}
=
\frac{1}{2}R^{\alpha}_{\fant{a}\beta\mu\nu}\theta^{\mu\nu}
=
d\omega^{\alpha}_{\fant{a}\beta}+\omega^{\alpha}_{\fant{a}\mu}\wdg \omega^{\mu}_{\fant{a}\beta},
\ee
where $R^{\alpha}_{\fant{a}\beta\mu\nu}$ are the components of the Riemann tensor in the orthonormal frame.
The contractions of the forms and tensor-valued forms are defined with the help of the contraction operator
$i_{e_\alpha}\equiv i_\alpha$ where $e_\alpha$ is the basis frame  fields that are
metric dual to the basis 1-forms: $i_\alpha\theta^\beta=\delta^{\beta}_{\alpha}$. The symbol $*$ corresponds to the Hodge dual operator that defines an inner product for two $p$-forms
in terms of the metric tensor. In terms of the Hodge dual, the oriented volume element reads
$*1=\sqrt{|g|}dx^{0}\wdg\cdots\wdg dx^{3}=\frac{1}{4!}\epsilon_{\alpha\beta\mu\nu}\theta^{\alpha\beta\mu\nu}$ in four spacetime dimensions where
$\epsilon_{\alpha\beta\mu\nu}$ is a completely antisymmetric permutation symbol. Ricci 1-forms can be defined as $R^{\alpha}\equiv i_\beta\Omega^{\beta\alpha}=R^{\beta\alpha}_{\fant{aa}\beta\mu}\theta^\mu$, whereas the scalar curvature $R$ can be written as $R=i_\alpha R^\alpha$. All the other geometrical objects will be introduced as they are required in terms of the quantities defined here.

The gravitational Lagrangians  will be assumed to depend on the set of basic gravitational variables $\{\theta^\alpha\}$ and $\{\omega^{\alpha}_{\fant{a}\beta}\}$ as well as their exterior derivatives, namely $d\theta^\alpha$ and $d\omega^{\alpha}_{\fant{a}\beta}$, through the scalars constructed from the contractions of the curvature tensor.
In a more general framework, minimal coupling of matter fields requires that the matter Lagrangians
involve $\theta^\alpha$ and $\omega^{\alpha}_{\fant{a}\beta}$
but not  $d\theta^\alpha$ and $d\omega^{\alpha}_{\fant{a}\beta}$ \cite{kopczynsky}.  Local Lorentz invariance of the gravitational Lagrangian forbids the explicit dependence of $\mathcal{L}$ on connection 1-forms and that the connection 1-forms  enter into  total Lagrangian via tensorial expressions and covariant exterior derivative. Moreover, in place of   $d\theta^\alpha$ and $d\omega^{\alpha}_{\fant{a}\beta}$ it is convenient to have the two forms $\Theta^\alpha$ and $\Omega^{\alpha}_{\fant{a}\beta}$ since general gravitational Lagrangians studied here are mainly based on scalars built on the curvature 2-forms.
In the explicit examples below, the Lagrangians will turn out  to depend on  particular nonminimally  coupled scalar fields as well.
For the sake of simplicity, the dependence on matter fields will be omitted. 
In deriving the metric field equations for gravitational models, the first order formalism where coframe $\{\theta^\alpha\}$ and
connection 1-forms $\{\omega^{\alpha}_{\fant{a}\beta}\}$ are regarded as the gravitational variables, will be used \cite{hehl,kopczynsky}. In this framework, the equations  for the pseudo-Riemannian metric are derived from the coframe variation of the Lagrangian subject to the constraint that the torsion 2-form $\Theta^\alpha=D\theta^\alpha$ vanishes. The metric compatibility of the connection 1-form $\omega_{\alpha\beta}+\omega_{\beta\alpha}=0$ is an algebraic constraint on the independent connection and it can simply be implemented into the total variational derivative by antisymmetrization
of the coefficients of $\delta\omega_{\alpha\beta}$. However, the torsion-free constraint,  namely $\Theta^\alpha=0$,   is a dynamical constraint and  can be imposed by extending the original Lagrangian density to include a Lagrange multiplier 2-form term $\mathcal{L}_{LM}=\lambda_\alpha\wdg \Theta^\alpha$ as
\be\label{gen-extended-lag}
\mathcal{L}_e[\theta^\alpha, \Omega^{\alpha}_{\fant{a}\beta},\Theta^\alpha, \lambda_\alpha]
=
\mathcal{L}[\theta^\alpha, \Omega^{\alpha}_{\fant{a}\beta}]
+
\lambda_\alpha\wdg \Theta^\alpha.
\ee
In the study of modified gravitational field equations relative to an orthonormal coframe, the use of derivatives of Lagrangian 4-forms with respect to $p$-form fields  are very convenient for the manipulation of the variational derivatives. Although it will not be made use of in the discussions below, it is possible to relate the derivatives of the volume form with  respect to a $p$-form to partial derivatives of appropriate scalars with respect to components of $p$-forms, we refer the reader to \cite{kopczynsky}.
By using the  variational derivatives of the curvature and torsion   2-forms
 \be
 \delta\Theta^\alpha
 =
 D\delta\theta^\alpha+\delta\omega^{\alpha}_{\fant{a}\beta}\wdg \theta^\beta,\qquad \delta\Omega^{\alpha}_{\fant{a}\beta}
 =
 D\delta\omega^{\alpha}_{\fant{a}\beta},
 \ee
the total variational derivative of extended Lagrangian density $\mathcal{L}_e$ with respect to the independent variables  can be found
to have the form
\begin{align}
\delta\mathcal{L}_e
&=
\delta\theta^\alpha\wdg
\left(
\frac{\partial \mathcal{L}_e}{\partial \theta^\alpha}+D\frac{\partial \mathcal{L}_e}{\partial \Theta^\alpha}
\right)
\nonumber
\\
&+
\delta\omega^{\alpha\beta}\wdg
\left[
D\frac{\partial \mathcal{L}_e}{\partial \Omega^{\alpha\beta}}
-
\frac{1}{2}
\left(
\theta_\alpha\wdg\frac{\partial \mathcal{L}_e}{\partial \Theta_\beta}
-
\theta_\beta\wdg\frac{\partial \mathcal{L}_e}{\partial \Theta_\alpha}
\right)
\right]
\nonumber
\\
&+
\delta\lambda^\alpha\wdg \frac{\partial \mathcal{L}_e}{\partial \lambda^\alpha}
+
d\left(
\delta\theta_\alpha\wdg\frac{\partial \mathcal{L}_e}{\partial \Theta^\beta}
+
\delta\omega^{\alpha\beta}\wdg
\frac{\partial \mathcal{L}_e}{\partial \Omega^{\alpha\beta}}
\right).\label{general-variation-explicit}
\end{align}
The definition of the derivative of the Lagrangian volume 4-form with respect to a $p$-form has been used with the usual partial derivative symbol  as is customarily done \cite{kopczynsky,obukhov} in the literature. For the explicit expression for $\mathcal{L}_e$ in (\ref{gen-extended-lag}), after evaluating some of the derivatives, the general expression (\ref{general-variation-explicit}) then takes the form
\begin{align}
\delta\mathcal{L}_e
&=
\delta\theta^\alpha\wdg *E_\alpha
\nonumber
\\
&+
\delta\omega_{\alpha\beta}\wdg
\left[
\Pi^{\alpha\beta}
-
\tfrac{1}{2}
\left(
\theta^\alpha\wdg\lambda^\beta
-
\theta^\beta\wdg\lambda^\alpha
\right)
\right]
\nonumber
\\
&+
\delta\lambda^\alpha\wdg \Theta_\alpha
+
d\left(
\delta\theta_\alpha\wdg\lambda^\alpha
+
\delta\omega_{\alpha\beta}\wdg*X^{\alpha\beta}
\right), \label{general-variation-exp}
\end{align}
where the following auxiliary tensor-valued forms  are defined in terms of the  derivatives that come out  in the variational derivative with respect to
gravitational variables
\be\label{aux-X-def}
\Pi^{\alpha\beta}
\equiv
D*X^{\alpha\beta}
\equiv
D\frac{\partial \mathcal{L}}{\partial \Omega^{\alpha\beta}}.
\ee
Assuming that the variations of the variables vanish on the boundary, the field equations for the connection 1-form  then
yields the algebraic relation
\be\label{conn-eqn-form1}
\Pi^{\alpha\beta}
=
\tfrac{1}{2}
\left(
\theta^\alpha\wdg\lambda^\beta
-
\theta^\beta\wdg\lambda^\alpha
\right).
\ee
The connection equations can be solved for the Lagrange multiplier  form $\lambda^\alpha$.
The Lagrange multiplier $(n-2)$-form $\lambda^\alpha$ is a vector-valued $2$-form while $\Pi^{\alpha\beta}$ is tensor-valued $3$-form  in four dimensions and they  have an equal number of components \cite{hehl}. The  equivalence can uniquely be expressed  by inverting (\ref{conn-eqn-form1}) to write $\lambda_\alpha$ in terms of $\Pi^{\alpha\beta}$ as
\be\label{general-expression-lag-mult}
\lambda^\beta
=
2i_\alpha\Pi^{\alpha\beta}-\tfrac{1}{2}\theta^\beta\wdg i_\mu i_\nu \Pi^{\mu\nu}.
\ee
This  formula can be obtained by calculating contractions of Eq. (\ref{conn-eqn-form1}).  The right-hand side is to be calculated subject to the vanishing torsion constraint $\Theta^\alpha=0$, which results from the variational derivative of the extended Lagrangian with respect to the Lagrange multiplier $2$-forms. As will be apparent with the explicit calculations in the subsequent sections, the auxiliary tensor-valued forms $X^{\alpha\beta}$ and $\Pi^{\alpha\beta}$ are also quietly suitable  in the construction  of a scalar-tensor equivalent Lagrangian for a given modified gravitational Lagrangian as well.

Finally, returning to the coframe equations,  in the general case, the metric equations are derived from the coframe variational derivative in the form
\be
\frac{\delta\mathcal{L}_e}{\delta\theta^\alpha}\equiv *E^\alpha=\frac{\partial\mathcal{L}}{\partial\theta^\alpha}+D\lambda^\alpha=0.
\ee
The Lagrange multipliers $\lambda_\alpha$, which are to be calculated by using (\ref{general-expression-lag-mult}) for the vanishing torsion constraint, also contribute to the metric equations.
The 1-form $E_\alpha=E_{\alpha\beta}\theta^\beta$ in the context of the discussion below, can be regarded as a generalization of  the  Einstein 1-form $G_{\alpha}=G_{\alpha\beta}\theta^\beta$. The tensor
$E_{\alpha\beta}$ is symmetric in its indices as a result of the local Lorentz invariance, that is, $\theta^\beta\wdg *E^\alpha-\theta^\alpha\wdg *E^\beta=0$, and it is covariantly constant as a result of the general coordinate invariance of the Lagrangian, i. e.,  $D*E^\alpha=0$ as in the case of the Einstein-Hilbert action.

The first order formalism naturally allows one to consider the connection as an independent gravitational variable,  as in the so-called Palatini approach, by dropping the constraint on the connection, i.e., by setting $\lambda_\alpha=0$. As a consequence, both the Palatini and the metric variational procedures yield the same metric equations only in the case $\lambda_\alpha=0$ \cite{kichenassamy}. The prime example of this exceptional case is the well-known Einstein-Hilbert action. The discussion  of the explicit examples below will   involve the pseudo-Riemannian case  by constraining the connection to be Levi-Civita. The field equations derived from the Palatini variational method will also be considered for $f(R)$ and for some curvature-squared gravitational  Lagrangian forms in connection with the construction of ST equivalence.

\section{ST equivalents for $f(R)$ theories}

In this section, in a streamlined fashion, the well-known equivalence between $f(R)$ theories and ST theories is presented
in the Riemannian case as well as the non-Riemannian geometries with either torsion or nonmetricity  of  $f(R)$ models relative to an orthonormal coframe with appropriate constraints. These $f(R)$ models and their ST equivalence have recently been studied in \cite{sotiriou2}.
The use of constrained first order formalism facilitates the study of all these subtheories in a unified manner and ST equivalents can be introduced in a unified manner as well.

An additional advantage of the presentation  in this framework, as indicated also in \cite{starobinski-shapiro},  is that it allows one  to further generalize such an equivalence for modified gravitational actions involving more general curvature invariants at the level of the field equations. In the subsequent section, the results will be generalized to modified gravitational actions based on quadratic curvature scalars. The  modified gravitational models will be considered without introducing  any matter field coupling.

\subsection{$f(R)$ model in the  Riemannian context}

It is well known that the field equations 
that follow from the modified gravitational action
\be\label{f(R)-action}
\mathcal{L}=f(R)*1
\ee
can be cast into equivalent Brans-Dicke-type scalar-tensor theory with a potential term. This is usually achieved by rewriting (\ref{f(R)-action}) as
\be
\mathcal{L}=[f(\chi)+f'(\chi)(\chi-R)]*1
\ee
using an auxiliary field $\chi$ where the prime denotes differentiation with respect to $\chi$. Under the condition $f''(\chi)\neq 0$, the field redefinition $f(\chi)$ then immediately yields the scalar-tensor equivalent for (\ref{f(R)-action}) see, e.g., \cite{faraoni,clifton-phys-rep}.

However, in defining scalar-tensor  equivalents below, a slightly different approach will be adopted by making use of the explicit expressions for the variational derivatives relative to an orthonormal coframe. Thus,  in order to derive the ST equivalent to the Lagrangian (\ref{f(R)-action}), it is convenient to begin with the variational derivative of the Lagrangian density (\ref{f(R)-action}) without considering the constraint term. Later, in deriving the corresponding field equations, the Lagrangian will be extended by the constraint term $\mathcal{L}_{LM}=\lambda_\alpha\wdg \Theta^\alpha$.
Explicitly, in the first order formalism,  the variational derivative  of (\ref{f(R)-action}) can be written in the form \cite{baykal-delice}
\be\label{f(R)-var}
\delta\mathcal{L}
=
(\delta f)*1+f\delta*1
=
f'\delta(R*1)+(f-Rf')\delta*1
\ee
without the need for further evaluation of the variations $\delta(R*1)$ and $\delta*1$.

First, an important observation that later will be made essential use of is that the expression (\ref{f(R)-var}) for  the variational derivative,  in fact, holds for any scalar built out of other curvature scalars  in the place of the simplest scalar $R$. Thus, (\ref{f(R)-var}) turns out to have  enough generality  that allows the extension of ST equivalence above.

Second, the coefficient of the variational derivative in the first term indicates  that the function $f'\equiv\frac{df}{dR}$ acts like a nonminimally coupled scalar field similar to the Brans-Dicke scalar field. The convenient field redefinitions that allow one to cast the theory into a ST-type theory
can in fact be found, without carrying out the variational derivatives further. The expression of the second term on the right-hand side of (\ref{f(R)-var}) is particularly useful in this regard.

On the other hand,  consider the following action
\be\label{st-eqivalent-action}
\mathcal{L}_{ST}[\phi, \theta^\alpha, \Omega^{\alpha}_{\fant{a}\beta}]
=
\phi R*1-V(\phi)*1,
\ee
with a typical scalar field nonminimally coupled  to gravity and having a potential term, which is to be made precise below, with  no accompanying kinetic term in contrast to the standard Brans-Dicke type ST Lagrangian. Note at this point that, although the scalar field has no kinetic term, as will also be shown below, the metric field equations render $\phi$ a dynamical field as a consequence of the nonminimal coupling. The total variational derivative of (\ref{st-eqivalent-action}) can easily be calculated to be
\be\label{st-var}
\delta\mathcal{L}_{ST}
=
\phi \delta(R*1)-V(\phi)\delta*1+ \delta\phi\left(R-\frac{dV}{d\phi}\right)*1.
\ee
The first term on the right-hand side of (\ref{st-var}) contributes to the  coframe as well as to the connection equations  whereas the
 second term contributes to the coframe equations.  Moreover, at this stage in deriving an ST equivalent Lagrangian, the explicit form of these  variational  derivatives is not needed. Finally, the last term on the right-hand side of the variational derivative (\ref{st-var}) 
 yields the scalar  field equation as
\be
R=\frac{dV}{d\phi}.
\ee
This  constraint  is identically satisfied  if the potential $V(\phi)$ is chosen to be the Legendre transform of $f(R)$, that is,
\be
V(\phi)
\equiv
\phi R-f(R),
\ee
where it is assumed that $f''\neq0$ and thus the definition $\frac{df}{dR}=f'(R)\equiv\phi$ is invertible so that one at least locally has $R=R(\phi)$. Therefore, the  potential $V(\phi)$ determined by  the explicit form of the function $f(R)$ and  the identification $f'\equiv\phi$ brings the modified action (\ref{f(R)-action})  to its ST equivalent given  in (\ref{st-eqivalent-action}). Consequently, it is possible to formulate the field equations using   either $\phi$ or $f'$. Thus, the construction of the dynamical equivalent model amounts to employing Legendre transform of $f(R)$ with the scalar field   $f'=\phi$. The ST equivalence leads to second order equations in the metric.
 In addition, it introduces a nonminimally coupled scalar field which satisfies some second order equations.

Subsequently, one  includes the constraint  that the torsion vanishes by a Lagrange multiplier term
$\mathcal{L}_{LM}=\lambda_\alpha\wdg \Theta^\alpha$ giving the extended Lagrangian density
\be
\mathcal{L}_e[f',\theta^\alpha,\Omega^{\alpha}_{\fant{a}\beta},\Theta^\alpha, \lambda_\alpha]
\equiv
\mathcal{L}[f',\theta^\alpha,\Omega^{\alpha}_{\fant{a}\beta}]
+
\mathcal{L}_{LM}.
\ee
and the metric equations then follow from the variational derivative of the  extended Lagrangian with respect to coframe 1-forms.
The total variational derivative is
\begin{align}
\delta\mathcal{L}_e
&=
\delta\theta_\alpha\wdg (f'\Omega_{\mu\nu}\wdg*\theta^{\alpha\mu\nu}+D\lambda^\alpha)
\nonumber
\\
&+
\delta\omega_{\alpha\beta}\wdg\left[D*f'\theta^{\alpha\beta}-\tfrac{1}{2}(\theta^\alpha\wdg \lambda^\beta-\theta^\beta\wdg \lambda^\alpha)\right]
\nonumber\\
&+
\delta\lambda_\alpha\wdg \Theta^\alpha
+d\left(
\delta\theta_\alpha\wdg\lambda^\alpha
+
\delta\omega_{\alpha\beta}\wdg*X^{\alpha\beta}
\right).
\label{f(R)-variation-evaluated}
\end{align}
In the present case, the explicit form of the auxiliary forms are $X^{\alpha\beta}=f'\theta^{\alpha\beta}$ and $\Pi^{\alpha\beta}=D*X^{\alpha\beta}$ which are to be calculated subject to the vanishing torsion constraint. Subsequently, the resulting expression  is used to express $\lambda^\alpha$ in terms of the gravitational variables with the help of (\ref{general-expression-lag-mult}). The metric field equations $\tfrac{1}{2}*E^\alpha=0$ that follow from coframe variation then take the form
\be\label{BD-type-eqn}
-f'*G^\alpha+\tfrac{1}{2}(f-Rf')*\theta^\alpha+D*(df'\wdg \theta^\alpha)=0.
\ee
Equivalently, in terms of the Legendre transform of $f(R)$, the field equations can be rewritten in terms of the scalar field $f'\equiv \phi$ as
\be\label{ohanlon-eqns}
-\phi*G^\alpha+D*(d\phi\wdg \theta^\alpha)-\tfrac{1}{2}V(\phi)*\theta^\alpha=0.
\ee
Although the field equations (\ref{ohanlon-eqns}) have resemblance with the Brans-Dicke (BD) field equations (cf., for example, the form of the Brans-Dicke equations given in \cite{baykal-delice-bd})
they do not involve the Brans-Dicke parameter $\omega$ as in the original BD Lagrangian. Such a theory (without the potential term for the scalar field) was introduced in connection with gravity theory with a Yukawa-type term in the Newtonian limit \cite{ohanlon}. On the other hand, on the basis of  ST equivalents of such theories, the value of the BD parameter $\omega=0$ implies that such theories are ruled out by solar system tests \cite{solar-test-chiba}.

The dynamical equations for the scalar field can be found  by tracing the metric equations and  one obtains the equation
\be\label{phi-eqn1}
\Delta\phi+U(\phi)=0,
\ee
where
\be
U(\phi)\equiv\frac{1}{(n-1)}\left[\phi\frac{dV}{d\phi}-(n-2)V(\phi)\right]
\ee
is introduced and the Laplace-Beltrami operator $\Delta$ can be defined  in terms
of the exterior differential and the codifferential $*d^\dagger=(-1)^pd*$ as $\Delta=d^\dagger d+dd^\dagger$ acting on $p$-forms.

In the Riemannian case discussed in this subsection, the Riemannian  connection is a quantity  derived from the metric. On the other hand, the modified gravitational Lagrangians of $f(R)$ form are mathematically simple enough to accommodate independent connection with torsion and/or nonmetricity. In such models the torsion and/or nonmetricity induces minimal coupling terms for the scalar field in the  scalar-tensor  type gravity \cite{dereli-tucker-non-riem}.
Consequently, the ST equivalents for the $f(R)$ models with torsion and/or nonmetricity are quite similar to the Riemannian case \cite{sotiriou-torsion}. The ST equivalents for these models are studied in some detail in the next two subsections using the first order formalism formulas discussed above.

\subsection{Riemann-Cartan type $f(R)$ model}

Although both the  Palatini and the metric variational approaches  yield the same equations   for the metric for the Einstein-Hilbert action,
for the modified actions of the form $f(R)*1$, they lead to distinct field equations. This subsection is devoted to  the $f(R)$ model  for which the connection is assumed to be metric compatible but have a nonvanishing torsion. The field equations can easily be obtained from those of the metric case.
Explicitly, by setting $\lambda^\alpha=0$  in (\ref{f(R)-variation-evaluated}), $R$ is assumed to be scalar curvature corresponding to the non-Riemannian connection $\Gamma^{\alpha}_{\fant{a}\beta}$. The connection equations $\Pi^{\alpha\beta}=0$ can be written as
\be\label{algebraic-torsion}
D(\Gamma) *f'\theta^{\alpha\beta}
=
df'\wdg *\theta^{\alpha\beta}+\Theta^\mu\wdg f'*\theta^{\alpha\beta}_{\fant{ad}\mu}
=0,
\ee
where $D(\Gamma)$ is now the covariant exterior derivative with respect to a connection form $\Gamma^\alpha_{\fant{a}\beta}$.
(\ref{algebraic-torsion}) is an algebraic equation for torsion  2-form and it can uniquely be solved for torsion as
\be\label{algebraic-torsion2}
\Theta^\mu
=
-\frac{1}{n-2}\theta^\mu\wdg d\ln f',
\ee
where $n$ denotes the  number of dimensions. Similarly, the coframe variations then take the form
\be\label{RC-metric-eqn}
-f'*G^\alpha(\Gamma)-\tfrac{1}{2}(R(\Gamma)f'-f)*\theta^\alpha=0,
\ee
where $*G^\alpha(\Gamma)$ is the Einstein form corresponding to the curvature of the connection $\Gamma$.
It is convenient to work with the ST equivalent which can be introduced  in the same way as before by  defining $\phi \equiv f'$ with the Legendre transformation   $V(\phi)\equiv R(\Gamma)f'-f$.

It is possible to rewrite (\ref{RC-metric-eqn}) in terms of Riemannian quantities  as follows. In order to do so one uses
the well-known fact that a general connection $\Gamma^{\alpha}_{\fant{a}\beta}$ can be decomposed into  the torsion and the nonmetricity parts
 in addition to a Riemannian part $\omega^{\alpha}_{\fant{a}\beta}$ as
\be\label{general-connection-decomposition}
\Gamma^{\alpha}_{\fant{a}\beta}
=
\omega^{\alpha}_{\fant{a}\beta}
+
K^{\alpha}_{\fant{a}\beta}
-
(i^\alpha Q_{\beta\mu}-i_\beta Q^\alpha _{\fant{a}\mu})\theta^\mu+Q^{\alpha}_{\fant{a}\beta},
\ee
where $K^{\alpha}_{\fant{a}\beta}$ is the contorsion 1-form defined in terms of the torsion  2-form as $\Theta^\alpha=K^{\alpha}_{\fant{a}\beta}\wdg\theta^\beta$
and  $Q_{\alpha\beta}$ is the nonmetricity 1-form defined by
\be
Q_{\alpha\beta}=-\tfrac{1}{2}D\eta_{\alpha\beta}=\tfrac{1}{2}(\Gamma_{\alpha\beta}+\Gamma_{\beta\alpha}),
\ee
and $\omega_{\alpha\beta}$ is the Riemannian connection satisfying $\omega_{\alpha\beta}+\omega_{\beta\alpha}=0$ and $d\theta^\alpha+\omega^\alpha_{\fant{a}\beta}\wdg\theta^\beta=0$
\cite{hehl,dereli-tucker-non-riem}.
Explicitly, in the particular case of a Riemann-Cartan type $f(R)$ model, by making use of (\ref{algebraic-torsion2})  one finds that
\be
K^{\alpha}_{\fant{a}\beta}
=
-
i^\alpha  (d\ln \phi)\theta_\beta+ i_\beta (d\ln \phi)\theta^\alpha,
\ee
and consequently, using this in (\ref{general-connection-decomposition}), it is possible to decompose the connection form
$\Gamma^\alpha_{\fant{a}\beta}$ into the Riemannian $\omega^\alpha_{\fant{a}\beta}$ and non-Riemannian  parts as
\be\label{f(R)-connections-relation}
\Gamma^\alpha_{\fant{a}\beta}
=
\omega^\alpha_{\fant{a}\beta}
-
i^\alpha  (d\ln \phi)\theta_\beta+ i_\beta (d\ln \phi)\theta^\alpha.
\ee
On the other hand, note that (\ref{f(R)-connections-relation}) is precisely the conformal transformation law  for the connection forms $\omega^\alpha_{\fant{a}\beta}$ under the conformal transformations of the metric $g\mapsto (\ln \phi)^2g$ in terms of the scalar field $\phi=f'$.  Consequently, in a Riemann-Cartan type $f(R)$ theory, which leads to an algebraic torsion, the decomposition (\ref{f(R)-connections-relation}) allows one to  write (\ref{RC-metric-eqn}) in  terms of pseudo-Riemannian quantities and a  scalar field as
\begin{align}
&-
\phi*G^\alpha(\omega)
+
D(\omega)*(d\phi\wdg \theta^\alpha)
-
\tfrac{1}{2}V(\phi)*\theta^\alpha
\nonumber
\\
&\quad  \quad \quad \quad \quad +
\left(\frac{n-1}{n-2}\right)\frac{1}{\phi}
*T^\alpha[\phi]
=0,
\label{RC-metric-eqn2}
\end{align}
where all quantities are pseudo-Riemannian and $T^\alpha[\phi]$ is the energy-momentum 3-form  for the scalar field defined by
\be\label{scalar-en-mom}
*T^\alpha[\phi]
=
-\tfrac{1}{2}\left\{(i^\alpha d\phi)*d\phi
+
d\phi\wdg i^\alpha*d\phi
\right\}.
\ee
Consequently, compared to the  equations (\ref{ohanlon-eqns}) of the metric case, the additional assumption of a nonvanishing torsion further  induces  scalar field  gravitational couplings reflected by the last term in (\ref{RC-metric-eqn}).
In contrast to the Einstein-Hilbert action, for the  general case $f(R)\neq R$, the Palatini variational procedure yields metric equations that are  not equivalent to those obtained by the metric variational procedure by construction \cite{faraoni,clifton-phys-rep}.
The geodesics defined by $\Gamma^{\alpha}_{\fant{a}\beta}$ will clearly be different from those of $\omega^{\alpha}_{\fant{a}\beta}$.

\subsection{$f(R)$ model with nonmetricity}

$f(R)$ model are simple yet rich enough to allow independent connection $\Gamma$ to have nonmetricity but vanishing torsion as well.
This subcase can be obtained from the Riemannian case by simply assuming  an algebraic nonmetricity constraint while keeping the zero torsion constraint.
In this case one has to pay attention to the fact that the raising and lowering of the indices with $\eta^{\alpha\beta}$
and $\eta_{\alpha\beta}$ do not commute with covariant exterior derivative $D(\Gamma)$.

Explicitly, the coframe equations  have formally the same as those
of the Riemannian  case, which takes the form
\be\label{coframe-eqn-nonmetricity-case}
-f'*G^\alpha(\Gamma)-\tfrac{1}{2}(R(\Gamma)f'-f)*\theta^\alpha+D(\Gamma)\lambda^\alpha=0,
\ee
where the Einstein form $*G^\alpha(\Gamma)$ and the scalar curvature $R(\Gamma)$ are to be now computed from curvature of the connection $\Gamma^{\alpha}_{\fant{a}\beta}$ with nonmetricity.
The equations for the connection  1-form ${\delta \mathcal{L}}/{\delta \Gamma^{\alpha}_{\fant{a}\beta}}=0$ accordingly take the form
\be
D(\Gamma)[f'*(\theta_{\alpha}\wdg \theta^\beta)]-\lambda_\alpha\wdg \theta^\beta=0,
\ee
subject to the vanishing torsion constraint $\Theta^\alpha=0$.
These equations can explicitly be written as
\be\label{connection-eqn-nonmetricity-case}
(df'\eta_{\alpha\mu}-\tfrac{1}{2}f'Q_{\alpha\mu})\wdg *\theta^{\mu\beta}+f'\eta_{\alpha\mu}\Theta_\nu\wdg *\theta^{\alpha\beta\nu}
-\theta_{\alpha}\wdg \lambda^\beta=0,
\ee
in terms of nonmetricity and torsion forms.
The connection equations in this case can be regarded  as an equation for both nonmetricity 1-form and the Lagrange multiplier 2-form.
The second term on the left-hand side is to be dropped because of the constraint $\Theta^\alpha=0$.
(\ref{connection-eqn-nonmetricity-case}) then admits the following simple solution for nonmetricity 1-form and the Lagrange multiplier 2-form.
If one assumes that
\be\label{nonmetricity-sol1}
\eta_{\alpha\mu}df'-\tfrac{1}{2}f'Q_{\alpha\mu}=0,
\ee
is satisfied by the nonmetricity 1-form  then one has $\lambda^\beta=0$.
The nonmetricity 1-form (\ref{nonmetricity-sol1})   has the same diagonal  elements and therefore is proportional to the Weyl 1-form
defined as the trace part of a general nonmetricity 1-form \cite{hehl}.

As in the Riemann-Cartan subcase, the connection $\Gamma^{\alpha}_{\fant{a}\beta}$ having nonmetricity
can be decomposed into Riemannian in addition to non-Riemannian parts as
\be\label{nonmetricity-connection-decomp}
\Gamma^{\alpha}_{\fant{a}\beta}
=
\omega^{\alpha}_{\fant{a}\beta}
-
2i^\alpha(d\ln f') \theta_{\beta}+2i_\beta(d\ln f') \theta^\alpha+2\delta^{\alpha}_{\beta}d\ln f'.
\ee
Thus, the $f(R)$ model also allows a connection with nonmetricity determined by the  gradient of the scalar $df'$.
The decomposition (\ref{nonmetricity-connection-decomp}) also allows one to write the coframe equations (\ref{coframe-eqn-nonmetricity-case}) in terms of Riemannian quantities corresponding to $\omega^{\alpha}_{\fant{a}\beta}$ and the scalar field $\phi\equiv f'$. One finds that (\ref{coframe-eqn-nonmetricity-case})
can be rewritten as
\begin{align}
&-
\phi*G^\alpha(\omega)
+
(\tfrac{n}{2}-1)D(\omega)*(d\phi\wdg \theta^\alpha)
-
\tfrac{1}{2}V(\phi)*\theta^\alpha
\nonumber
\\
&+
\frac{(2-n)(n-1)}{2\phi}
\left[
(i^\mu d\phi)\wdg*d\phi-\frac{n-7}{n-1}d\phi\wdg i^\mu *d\phi
\right]
\nonumber \\
&=0. \label{nonmetricity-metric-eqn}
\end{align}
As in the case of the Riemann-Cartan type $f(R)$ model, the nonmetricity gradient introduces an additional gravitational coupling term of  scalar field to gravity.
However, the induced scalar coupling terms explicitly depend on the number of the dimensions and in particular for the number of dimension $n=4$,  the expression
reduces to
\be\label{nonmetricity-metric-eqn-4d}
-
\phi*G^\alpha(\omega)
+
D(\omega)*(d\phi\wdg \theta^\alpha)
-
\tfrac{1}{2}V(\phi)*\theta^\alpha
+
\frac{3}{\phi}*T^{\alpha}[\phi]
=
0
\ee
where $*T^{\alpha}[\phi]$ is defined as in (\ref{scalar-en-mom}).

It is interesting to note that it is possible to constrain nonmetricity form by introducing the constraint
term
\be\label{nonmetricity-C-constraint}
\rho^{\beta}_{\fant{a}\alpha}\wdg (Q^{\alpha}_{\fant{a}\beta}-\delta^{\alpha}_{\beta}d h(R)),
\ee
where $h(R)$ is a given function of the scalar curvature and $\rho^{\beta}_{\fant{a}\alpha}$ is symmetric Lagrange multiplier $(n-1)$-form.
The variational derivative of the $f(R)$ model with the constraint (\ref{nonmetricity-C-constraint}) is, however, technically  more involved  than the case discussed.
At this point, it is convenient to note that the constrained  first order formalism encompasses the recent C-theories which extrapolate and interpolate the metric and the Palatini methods \cite{amendola}. In the general C-theory framework, independent connection has nonmetricity characterized by a (co)vector
and it is derived from a  metric conformally related to the independent metric depending only on the scalar curvature of the connection.
On the other hand, in the first order formalism one proceeds in the opposite direction starting with a  prescribed nonmetricity on
independent connection \cite{koivisto13}.

The two simple  non-Riemannian cases discussed above therefore imply that  a general  $f(R)$ model can be formulated in the framework of constrained first order formalism to have prescribed torsion and/or nonmetricity and these non-Riemannian models can be cast in a Riemannian form with the non-Riemannian quantities inducing further gravitational couplings of the scalar fields in the corresponding  ST equivalents.

With regard to the ST equivalence, the generalization of the above discussion to the case with general curvature invariant in the place of the scalar curvature $R$ will explicitly studied below. The connection equations for generic  $f(R)$ models be involve the term $\partial \mathcal{L}/\partial \Omega_{\alpha\beta}=f'*\theta^{\alpha\beta}$ and lead to an algebraic equation for torsion and therefore the torsion does not propagate.
On the other hand, the connection equations are more complicated since $\partial \mathcal{L}/\partial \Omega_{\alpha\beta}=f'*\theta^{\alpha\beta}$ is more complicated than those of corresponding  $f(R)$  models discussed above. Therefore, it is technically more difficult to treat   more complicated gravitational Lagrangians  in metric  and Palatini formulations within the same framework and the corresponding ST equivalents in these distinct cases will have quite distinct features as well.

In the next section, a general form of the field equations for a modified Lagrangian based on an arbitrary curvature invariant and the construction of an ST equivalent is presented  and the dynamical degree of freedom carried by the resulting scalar field  is scrutinized.

\section{General scalar-tensor equivalence for an arbitrary curvature invariant}

A salient feature of the  above discussion  is that the  study of the field equations for $f(R)$ theory can be carried out using an arbitrary scalar built out of the curvature tensor, say $P$. The above analysis  can be generalized  to relate the equations of motion for $\mathcal{L}_P=P*1$, which can be written in the above notation as $*E^\alpha_P=0$, to those of    $\mathcal{L}_{mod.P}=f(P)*1$ without specifying the explicit form of the scalar $P$, except it is assumed that
it is constructed out of an arbitrary contraction of Riemann tensor. It is also not necessary to give the explicit form of the field equations derived from the Lagrangian form $P*1.$

Consider, now  the variational derivative of  $\mathcal{L}_{mod. P}$ for a given  function $f(P)$. As in the  $f(R)$ case, it is convenient to put it into the following preliminary form
\be\label{preliminary-gen1}
\delta\mathcal{L}_{mod. P}
=
f'\delta\mathcal{L}_P-\delta\theta^\alpha\wdg (Pf'-f(P))*\theta^\alpha.
\ee
Now, one can make use of the general expression (\ref{general-variation-explicit}) for $\delta\mathcal{L}_P$ in (\ref{preliminary-gen1}) to obtain the variational derivative of the extended Lagrangian $\mathcal{L}_{mod.P}+\mathcal{L}_{LM}$ in the form
\begin{align}
\delta &(\mathcal{L}_{mod.P}+\mathcal{L}_{LM})
\nonumber \\
&=
\delta\theta_\alpha\wdg \left[f'\frac{\partial \mathcal{L}_P}{\partial \theta_\alpha}+D\lambda^\alpha+(f-Pf')*\theta^\alpha\right]
\nonumber \\
&+
\delta\omega_{\alpha\beta}
\wdg
\left[
D\left(f'\frac{\partial\mathcal{L}_{P}}{\partial \Omega^{\alpha\beta}}\right)
-
\tfrac{1}{2}
\left(
\theta^\alpha\wdg\lambda^\beta
-
\theta^\beta\wdg\lambda^\alpha
\right)
\right]
\nonumber
\\
&+
\delta\lambda_\alpha\wdg \Theta^\alpha
+
d\left(
\delta\theta_\alpha\wdg\lambda^\alpha
+
\delta\omega_{\alpha\beta}\wdg
f'\frac{\partial\mathcal{L}_{P}}{\partial \Omega^{\alpha\beta}}
\right).
\end{align}
This result expresses the relation between the auxiliary forms corresponding to  the modified Lagrangian $\mathcal{L}_{mod. P}$
and the original Lagrangian $\mathcal{L}_{P}$ for the partial derivatives with respect to the gravitational variables. By considering (\ref{aux-X-def}),
it is convenient to define the auxiliary tensor valued forms as
\be
*X^{\alpha\beta}_m
=
f' *X^{\alpha\beta}_P,
\ee
and
\be\label{pi-decomp}
\Pi^{\alpha\beta}_m
=
D*X^{\alpha\beta}_m
=
f'\Pi^{\alpha\beta}_P+\tilde{\Pi}^{\alpha\beta},
\ee
in terms of the auxiliary forms $X^{\alpha\beta}_P$ and $\Pi^{\alpha\beta}_P$ of the  $\mathcal{L}_P$.
In Eq. (\ref{pi-decomp}), for convenience, $\tilde\Pi^{\alpha\beta}$   is defined to be
\be
\tilde{\Pi}^{\alpha\beta}\equiv df'\wdg *X^{\alpha\beta}_P.
\ee
These definitions  then lead to the following convenient split of the Lagrange multiplier form:
\be
\lambda^{\alpha}
=
f'\lambda^{\alpha}_P+\tilde{\lambda}^\alpha,
\ee
with the second term is given by
\be
\tilde{\lambda}^\alpha
=
2i_\beta\tilde{\Pi}^{\beta\alpha}+\tfrac{1}{2}\theta^\alpha\wdg i_\mu i_\nu \tilde{\Pi}^{\mu\nu}.
\ee
Finally, the coframe  equations for the modified Lagrangian $f(P)*1$ then take the following general form
\be\label{gen-f(P)-eqn}
f'*E_P^\alpha-(Pf'-f(P))*\theta^\alpha
+
df'\wdg \lambda^{\alpha}_P+D\tilde{\lambda}^\alpha
=0.
\ee
This equation can be considered as a slightly generalized form of the metric $f(R)$ equations (\ref{BD-type-eqn}).
If one takes  $P=R$ in the above formulas,  one then has $\lambda_R^\alpha=0$ identically with $\tilde{\lambda}^\alpha=4*(df'\wdg\theta^\alpha)$
and consequently (\ref{gen-f(P)-eqn})  reduces to (\ref{BD-type-eqn}).

An  equation for the scalar $f'$, as in the $f(R)$ case, can be  derived  by tracing the metric equations. The trace can  explicitly be written as
\be\label{scalar-eqn-general}
-
2di_\beta(df'\wdg*i_\alpha X^{\alpha\beta}_P)
-
df'\wdg \theta_\alpha\wdg \lambda^{\alpha}_P+[4f+(E-4P)f']*1=0
\ee
where $E\equiv E^{\alpha}_{\fant{a}\alpha}$
and $\lambda^{\alpha}_P$ is the Lagrange multiplier of the Lagrangian form $P*1$.  Thus,
the second term contains only the first order derivatives of  $f'$  whereas the first term contains the second order derivatives of $f'$.
Recalling the definition of the auxiliary form (\ref{aux-X-def}), (\ref{scalar-eqn-general}) implies that if the Lagrangian form $\mathcal{L}_P=P*1$ is
such  that
\be\label{dynamical-condition}
 *i_\alpha X^{\alpha\beta}_P
=
-\theta_\alpha\wdg\frac{\partial \mathcal{L}_P}{\partial \Omega_{\alpha\beta}}=0,
\ee
then the equations for the scalar $f'$ reduces to the algebraic equation $4f+(E-4P)f'=0$.
Note here that the second term in (\ref{scalar-eqn-general}) also involves $i_\alpha X^{\alpha\beta}_P$ since
 \be\label{case1-vanish}
 \theta_\alpha\wdg \lambda^\alpha_P=2i_\beta D*(i_\alpha X^{\alpha\beta}_P).
\ee
There are well-known cases where (\ref{case1-vanish}) is satisfied identically.
Consider the gravitational Lagrangians involving only the quadratic curvature expressions,
where $P$ is then quadratic in curvature components. In four dimensions,
there is only one  Lagrangian form  for which (\ref{case1-vanish}) vanishes identically
\be
\lambda^\alpha_P=2*C^\alpha,
\ee
where $*C^\alpha$ is the Cotton 2-form defined by $C^\alpha\equiv D(R^\alpha-\frac{1}{6}R\,\theta^\alpha)$ in four dimensions \cite{cotton}. In this case, it is easy to deduce that the corresponding quadratic curvature (QC) Lagrangian  is then of the form $\mathcal{L}_P=P*1=R^\alpha\wdg *R_\alpha-\frac{1}{3}R^2*1$, and is equivalent to a Weyl-squared Lagrangian up to a boundary term. This case will be discussed in a separate section below. For generic QC gravity, this conclusion can be extended to arbitrary dimensions simply by singling out the QC gravity Lagrangians whose field equations have second order trace. The QC Lagrangians having this particular property in arbitrary dimension $n\geq3$ have been  constructed in \cite{baykal}.

Any modified Lagrangian based on the general form $f(\mbox{Riemann}^2)$ has an ST equivalent with dynamical scalar field. This follows from the fact that,
with the definition $\Omega_{\alpha\beta}\wdg *\Omega^{\alpha\beta}\equiv K*1$ for Lagrangian form $f(K)*1$, the auxiliary form $X^{\alpha\beta}$ is
given by $\frac{df}{dK}*\Omega^{\alpha\beta}$ and its  contraction does not vanish identically.

A trivial case for which the scalar field is nondynamical is the case where $\lambda_P^\alpha=0$ identically. In  this case
$\lambda_P^\alpha=0$ follows from  $\Pi^{\alpha\beta}_P=D*X_P^{\alpha\beta}=0$ and  consequently the only nonvanishing contribution to the
modified Lagrange multiplier form comes from $\tilde{\Pi}^{\alpha\beta}=df'\wdg *X^{\alpha\beta}$,  which also renders $f'$ non-dynamical.
Recalling the expression  of the auxiliary form $*X_P^{\alpha\beta}$ defined in (\ref{aux-X-def}), the vanishing of Lagrange multiplier
\emph{in the metric case} requires
\be\label{requirement-GB}
D*X_P^{\alpha\beta}=D\frac{\partial\mathcal{L}_P}{\partial\Omega_{\alpha\beta}}=0,
\ee
subject to the condition $\Theta^{\alpha}=0$.  This requirement is identically satisfied if the partial derivative in (\ref{requirement-GB})  is, for example,  of the form $*\theta^{\alpha\beta}$ corresponding to Einstein-Hilbert action. The next complicated  example for (\ref{requirement-GB}) is satisfied when the partial derivative is of the form
\be\label{curvature-derivative-GB}
\frac{\partial\mathcal{L}_P}{\partial\Omega_{\alpha\beta}}
=
\Omega_{\mu\nu}\wdg*\theta^{\alpha\beta\mu\nu},
\ee
for which $D*X_P^{\alpha\beta}\equiv0$ as a consequence of the Bianchi identity, $D\Omega_{\mu\nu}\equiv0$, in addition to  the vanishing torsion constraint
$D\theta^\alpha=0$.
``Integrating" (\ref{curvature-derivative-GB}), one finds that $\mathcal{L}_P$ is the well-known Gauss-Bonnet Lagrangian density
\be
G*1\equiv\Omega_{\alpha\beta}\wdg \Omega_{\mu\nu}\wdg*\theta^{\alpha\beta\mu\nu}.
\ee
Consequently, for a modified gravitational Lagrangian density of the form $f(G)*1$,  the scalar tensor equivalence yields only a superficial field redefinition
for $f'$, since there is no dynamical degree of freedom resulting from the function $f'$.

The use of the first order formalism relative to an orthonormal coframe makes the construction of ST equivalents transparent for a given modified gravitational Lagrangian and allows one to write the field equations in a formally simplified form in the pseudo-Riemannian case.
The new insight gained through the study of the Lagrange multiplier term of the modified Lagrangian in terms of the original Lagrangian is that the equation for the scalar field in a generic ST equivalent is determined by the properties  of the field equations of the original Lagrangian and the dynamical status  of the scalar field depends on  the form of the original Lagrangian through the trace  of the derivative ${\partial \mathcal{L}_P}/{\partial \Omega_{\alpha\beta}}$.

Although it is only technically more involved, it is straightforward to extend the above discussion to the case where $f$ depends on an arbitrary number of  arguments. In particular, note that the form of the metric equations (\ref{gen-f(P)-eqn}), independent of the explicit  form of the scalar $P$, facilitates  the field redefinition $f'=\phi$ and thus it motivates a dynamically equivalent  multi-scalar-tensor  model for the modified gravitational Lagrangian densities of the form $\mathcal{L}=f(R, Q, P, K,\ldots)*1$ where $Q, P, K, \ldots$ are distinct scalars  in the components of the curvature tensor.
In the general case, the scalar $f'$ has nonminimal coupling and is dynamical. On the other hand, in the case where the scalar field is nondynamical,
the Legendre transform of $f$ becomes proportional to the trace of the equations that follow from $P*1$ and the field redefinition $f'\equiv \phi$ becomes superfluous.

As will also be studied to some extent in the examples below, the number of scalar fields in the ST equivalents can be less than the number of scalars on which a general $f$ depends. This is, in fact related to the property of the Legendre transform with constraints and consequently the number of scalar field equations can be less then the number of independent arguments of $f$. The constraints in the present context though are not related to the evolution of the gravitational field variables \cite{starobinski-shapiro}.

 In the next section, these considerations will be applied to the explicit cases where the scalars $Q, P, K$ are assumed to be some quadratic curvature scalars and  to their corresponding ST equivalents, which  are relevant to some popular cosmological models.

\section{Modified quadratic curvature Lagrangians}

In order to study the multi-ST equivalents for modified gravity models at the level of the  field equations, it is first convenient to study the field equations that follow from quadratic curvature models. Eventually, the field equations for the modified  models  will be formulated in terms of quadratic curvature gravity equations with nonminimally coupled scalar fields.

In four dimensions, there are three independent quadratic curvature scalars, namely $R^2*1, R_\alpha\wdg*R^\alpha$ and $ \Omega_{\alpha\beta}\wdg*\Omega^{\alpha\beta}$. Therefore the most general gravitational action density involving quadratic  invariants consists of linear combinations of these individual densities.
For what follows it is convenient to introduce the scalars $Q, P,K$ by the following definitions
\be\label{individual-qc-lag}
\begin{split}
\mathcal{L}_Q
&=
Q*1
\equiv
R^2*1,\\
\mathcal{L}_P
&=
P*1
\equiv
R^\alpha\wdg *R_\alpha,\\
\mathcal{L}_K
&=
K*1
\equiv
\Omega^{\alpha\beta}\wdg *\Omega_{\alpha\beta}.
\end{split}
\ee
In this section ST equivalents for the Lagrangian density of the form $f(Q, P, K)*1$ will explicitly be constructed and the corresponding field equations
will be derived in some generality.

As a result of the well-known fact that, in four spacetime dimensions, the variational  derivative of the Gauss-Bonnet term vanishes \cite{lovelock}, i.e., $\delta (\Omega_{\alpha\beta}\wdg \Omega_{\mu\nu}\wdg *\theta^{\alpha\beta\mu\nu})=0$, only two of the three Lagrangian densities  in (\ref{individual-qc-lag}) are functionally independent. However, intending the  modification of the quadratic curvature Lagrangian densities in the $f(R)$ spirit, it is convenient to  work with all three scalars. For further calculational details for the variational derivatives of the Lagrangians relative to an orthonormal coframe considered here, see \cite{baykal,baykal-delice,var-dereli-tucker}.

Since the field equations for any of the Lagrangians in (\ref{individual-qc-lag})  can be written in the same particular form, it is convenient to define the following total Lagrangian density,
\be\label{n>4-qc-lag}
\mathcal{L}_{qc}
=
a\,\mathcal{L}_Q+b\,\mathcal{L}_P+c\,\mathcal{L}_K,
\ee
where $a,b,c$ are arbitrary coupling constants and Latin subscripts $ Q, P, K$, for example,  indicate tensorial quantities belonging to the Lagrangians $\mathcal{L}_Q, \mathcal{L}_P, \mathcal{L}_K$ respectively.
Note that the scalar fields in the equivalent Lagrangians are not dynamical for every combination of the QC coupling constants  $a,b,c$.
For (\ref{n>4-qc-lag}), one has
\be\label{qc-en-mom}
\frac{\partial \mathcal{L}_{qc}}{\partial \theta^\mu}
=
\Omega_{\alpha\beta}\wdg i^\mu*X^{\alpha\beta}_t-i^\mu\mathcal{L}_{qc}
\equiv
*T^\mu_{qc},
\ee
where the auxiliary 2-form is explicitly given by
\begin{align}
\label{aux-X-n+d}
X_{qc}^{\alpha\beta}
&=
a\,X_Q^{\alpha\beta}
+
b\,X_P^{\alpha\beta}
+
c\,X_K^{\alpha\beta}
\\
&=
2a\,R\theta^{\alpha\beta}
+
b\,(\theta^{\alpha}\wdg R^\beta-\theta^\beta\wdg R^\alpha)
+
2c\,\Omega^{\alpha\beta}.\nonumber
\end{align}
Note that (\ref{qc-en-mom}) can be written in a linear sum of the form
\be
*T^\alpha_{qc}
=
a*T^\alpha_Q+b*T^\alpha_P+c*T^\alpha_K,
\ee
with the help of (\ref{aux-X-n+d}). The auxiliary  3-form $*T^\alpha_{qc}$ has  a mathematical structure that is formally analogous  to energy-momentum form of  a generic minimally coupled matter 2-form field $F=\tfrac{1}{2}F_{\alpha\beta}\theta^{\alpha\beta}$ expressed relative to an orthonormal coframe (see, e.g., \cite{thirring}). It is, for example, traceless (only in four dimensions): $T^{\alpha}_{{qc}\alpha}=0$. Thus, only the fourth order terms, the Lagrange multiplier terms, contribute to the trace of the metric equations.
$X_{qc}^{\alpha\beta}$ is used to calculate $\lambda^\alpha_{qc}$ making use of  $\Pi^{\alpha\beta}_{qc}=D*X_{qc}^{\alpha\beta}$
and  the Lagrange multiplier can similarly be written as the linear sum of multipliers  for individual QC components as
\begin{align}
\lambda_{qc}^\alpha
&=
a\lambda_Q^\alpha+b\lambda_P^\alpha+c\lambda_K^\alpha
\nonumber\\
&=
4a*(dR\wdg \theta^\alpha)
+
b*D(2R^\alpha+R\theta^\alpha)
\nonumber
\\
&
+
c(4i_\beta D*\Omega^{\beta\alpha}+\theta^\alpha\wdg i_\mu i_\nu D*\Omega^{\mu\nu}).
\end{align}
 Consequently, with the help of the general formulas provided above, the field equations for the general  quadratic curvature Lagrangian  (\ref{n>4-qc-lag})  can be written in the concise form
\be\label{4+d-field-eqn}
*E^\alpha_{qc}
=
D\lambda_{qc}^\alpha
+
*T^\alpha_{qc}
=0.
\ee
In the general case corresponding to (\ref{n>4-qc-lag}), the form of the field equations is dimension independent and the trace of the field equations  (\ref{4+d-field-eqn})  in $n\geq 3$ can be found as
\be
E_{qc}*1
=
-[4a(n-1)+nb+c]d*dR+(n-4)\mathcal{L}_{qc},
\ee
where $E_{qc}*1\equiv \theta_\alpha\wdg *E^\alpha_{qc}$ as before and in the present work we confine the study to $n=4$ dimensions.

After the discussion of the QC field equations in an appropriate form, now it is convenient to consider the  gravitational Lagrangians that depend on the quadratic curvature invariants $Q, P, K$  in the form
\be
\mathcal{L}_{mod.}
=
f(Q, P, K)*1.
\ee
Before studying modified gravitational field equations for specified functions $f$,  it is first convenient to derive the field equations in some generality.
As before, it is convenient to write the total variational derivative of the Lagrangian $\mathcal{L}_{mod.}$
in the form
\begin{align}
\delta\mathcal{L}_{mod.}
&=
f_Q\delta\mathcal{L}_Q+f_P\delta\mathcal{L}_P+f_K\delta\mathcal{L}_K
 \nonumber \\
&+ (f-Qf_Q-Pf_P-Kf_K)\delta*1.
\end{align}
The explicit form of the last term on the right-hand side, in this case, is the Legendre transform of the function $f(Q, P, K)$.
By introducing the scalar fields
\be \label{multiscalars}
\phi_1\equiv f_Q=\frac{\partial f}{\partial Q},\ \phi_2\equiv f_P=\frac{\partial f}{\partial P},\ \phi_3\equiv f_K=\frac{\partial f}{\partial K},
\ee
the variational derivative of the modified Lagrangian $\mathcal{L}_m$  becomes equivalent to that of the multi-scalar-tensor-type Lagrangian
\be\label{ST-eq-action}
\mathcal{L}_{ST}
=
\phi_1\mathcal{L}_Q+\phi_2\mathcal{L}_P+\phi_3\mathcal{L}_K-V(\phi_1,\phi_2,\phi_3)*1,
\ee
provided that the potential term $V(\phi_1,\phi_2,\phi_3)$ for the scalar fields is taken to be the Legendre transform of $f(Q, P, K)$, namely
\be\label{multi-scalar-pot}
V(\phi_1,\phi_2,\phi_3)
\equiv
Qf_Q+Pf_P+Kf_K-f(Q,P,K).
\ee
Note that the labels $Q, P, K$ stand for derivative only when they are written as a subscript of the function $f$ and in all other cases it is a discriminating label for the corresponding QC quantities.

Similar to the $f(R)$ theory case, the dependence of the multivariable function $f$ on a particular scalar introduces a corresponding scalar field by means of the Legendre transform. The number of scalar fields $\phi_k$ does in fact depend on the explicit form of the function $f$ and is equal to the rank of the Hessian matrix related to the Legendre transform of $f$. It is assumed that the function $f$  satisfies some regularity condition which ensures that the Legendre transformation is invertible. The Hessian matrix of the Legendre transformation $\{f_P, f_Q, f_K\}\mapsto \{\phi_1,\phi_2,\phi_3\} $ is  assumed to be of rank three \cite{starobinski-shapiro}. For the explicit examples discussed below, the relevant Hessian matrices  are of lower rank and consequently the general formulas for multi-ST equivalence simplify considerably.

 In terms of the new field variables defined above, the equations for the modified  gravitational Lagrangian can be written concisely in the form
\begin{widetext}
\begin{align}
&\phi_1 *E^\alpha_Q+\phi_2 *E^\alpha_P+\phi_3 *E^\alpha_K-V(\phi_1, \phi_2, \phi_3)*\theta^\alpha
+
2d\phi_1\wdg i_\beta D*X^{\alpha\beta}_Q
+
2d\phi_2\wdg i_\beta D*X^{\alpha\beta}_P
\nonumber\\
&
+
2d\phi_3\wdg i_\beta D*X^{\alpha\beta}_K
+
2D[(i_\beta d\phi_1)*X^{\alpha\beta}_Q+(i_\beta d\phi_2)*X^{\alpha\beta}_P+(i_\beta d\phi_3)*X^{\alpha\beta}_K]=0,
\label{general-modified-eqn}
\end{align}
\end{widetext}
relative to an orthonormal coframe \cite{baykal}. Equations (\ref{general-modified-eqn}) can be considered to be an extension of  the field equations (\ref{ohanlon-eqns}). These metric equations follow from their multi-ST equivalent actions (\ref{ST-eq-action}) which are called  ``dual"
to the original modified actions \cite{starobinski-shapiro}.

For generic quadratic curvature gravity, the tensor-valued auxiliary forms $X^{\alpha\beta}_{qc}$ above  involve the terms that are linear in the contraction of the curvature 2-form, Ricci 1-form  and scalar curvature and consequently, the terms containing $X^{\alpha\beta}$
 in (\ref{general-modified-eqn}) can be regarded as the interaction terms of the scalar fields $\phi_k$ with curvature for $k=1, 2, 3$.  These terms generalize the second term in (\ref{BD-type-eqn}) in the sense that the expression $f'\theta^{\alpha\beta}$ for the $f(R)$ model is replaced with the above $X^{\alpha\beta}$'s that are linear in curvature components.
The equations for the scalar fields then can be obtained by tracing the field equations as in the $f(R)$ case.
In particular, tracing the general equation
(\ref{general-modified-eqn}), it is easy to find that the scalar fields become nondynamical if the corresponding $X^{\alpha\beta}$ satisfies
$i_\alpha X^{\alpha\beta}=0$. This important subcase  will be studied in one of the case studies below.
However, note that, in contrast to the simpler $f(R)$ case, the metric equations for the ST equivalent model are still fourth order in metric components in this case as well.

The field equation (\ref{general-modified-eqn}) encompasses all the cases below except for the sixth order case where the generalization of $dR\wdg*dR\equiv S*1$ in the form $f(S)$ will also be considered.  However, because the  derivation of the field equations proceeds  in a slightly different way than other quadratic curvature Lagrangians, it  will be treated  separately    below. However,  it is easy to include the scalar $S$ in the above framework once the corresponding field equations are obtained. Various  examples that are of interest in applications of modified gravity ranging from  cosmology to the topics relevant to the quantum properties of black holes are studied below.

\subsection{Modified Ricci-squared  Lagrangian}

For a relatively simple application  of the use of the general result obtained in the previous section, consider the  case where $f=f(P)$ where $P$ is defined in (\ref{individual-qc-lag}).  By making use of (\ref{general-modified-eqn}), the field equations for the Lagrangian density $f(P)*1$ can immediately  be written down as
\begin{align}
\phi  *E^\alpha_P
-
V(\phi)*\theta^\alpha
+
2d\phi\wdg i_\beta D*X^{\alpha\beta}_P
\nonumber
\\
+
2D[(d\phi)_\beta*X^{\alpha\beta}_P]=0,
\label{f(P)-eqns}
\end{align}
where in this case, $\phi\equiv \frac{df}{dP}$, the potential is given by $V(\phi)=P\frac{df}{dP}-f(P)$ and the explicit form of the equations for
the Ricci-squared  Lagrangian density $P*1$ takes the form
\begin{align}
*E^\mu_P
=
D*D(2R^\mu+& R\theta^\mu)-\tfrac{1}{2}(i^\mu\Omega_{\alpha\beta})\wdg *X^{\alpha\beta}_P \nonumber
\\&
+\tfrac{1}{2}\Omega_{\alpha\beta}\wdg i^\mu*X^{\alpha\beta}_P=0,
\label{P-eqns}
\end{align}
with $
X^{\alpha\beta}_P
=
\theta^\alpha\wdg R^\beta-\theta^\beta\wdg R^\alpha.
$
Inserting the expression (\ref{P-eqns}) into (\ref{f(P)-eqns}) and then calculating the trace, after some algebra, one obtains
\begin{align}
\phi\, d*dR-4V(\phi)*1+2\,d\phi\wdg *dR\nonumber \\+2\,d[(d\phi)_\alpha*R^{\alpha}]+R\,d*d\phi=0.
\end{align}
The importance  of the result (\ref{f(P)-eqns})  becomes more pronounced when it is considered in connection with  the result that
the modified gravitational Lagrangian based  on $f(R, Q,P)=R+aQ+bP$ is equivalent to Einstein  gravity interacting with
additional fields \cite{magnano-grg,jakubiec}.

The field equations for the $f(P)$ Lagrangian in the first order formalism can easily be  found simply by  disregarding the term imposing the constraint $\Theta^\alpha=0$. In this case, the independent (metric compatible) connection and coframe equations respectively take the form
\begin{align}
D(\Gamma)*f'X^{\alpha\beta}_P=0,
\label{connection-eqn-Q-palatini}\\
f'*T^{\alpha}_P[\Omega_{\mu\nu}(\Gamma), X_P^{\mu\nu}]+(f-Pf')*\theta^\alpha=0,
\label{coframe-eqn-Q-palatini}
\end{align}
where $\Omega_{\mu\nu}(\Gamma)$ is the curvature 2-form corresponding to the  connection $\Gamma_{\alpha\beta}$
and $D(\Gamma)$ stands for the covariant exterior derivative corresponding to the connection $\Gamma_{\alpha\beta}$. In  the case $f(P)=P$,
assuming that the connection  $\Gamma_{\alpha\beta}$ is torsion free, for an Einstein manifold with the Ricci 1-form of the form
$R^\alpha=k\,\theta^\alpha$  for some nonzero constant $k$, the connection equations (\ref{connection-eqn-Q-palatini}) are satisfied identically
since $X_P^{\alpha\beta}$ simplifies to $X_P^{\alpha\beta}=2k\,\theta^{\alpha\beta}$.
Moreover, the auxiliary 3-form  $*T^{\alpha}_P[\Omega_{\mu\nu}(\Gamma), X_P^{\mu\nu}]$, in this case  takes the form
\be
*T^{\alpha}_P[\Omega_{\mu\nu}(\Gamma), X_P^{\mu\nu}]
=
k\,\Omega_{\mu\nu}(\Gamma)\wdg * \theta^{\alpha\mu\nu}
-
4k^2*\theta^\alpha,
\ee
and consequently, the coframe equations (\ref{coframe-eqn-Q-palatini}) become the familiar vacuum Einstein field equations with a cosmological constant
\be
-
2f'k*G^\alpha
-
f'4k^2*\theta^\alpha
+
(f-4k^2f')*\theta^\alpha
=0,
\ee
where $P=2k^2$ for which $f(P)$ and $f'(P)$ are also  constants.
These equations then constitute a certain  part of the theorem on the so-called \emph{universal} property of the Lagrangians depending  on the Ricci-squared  scalar \cite{borowiec}. Note however that in a metric theory, the generic quadratic curvature Lagrangians (as well as their modification in the $f(P)$ fashion) lead to fourth order equations,   and they admit  Einstein metric solutions for which the fourth order term  ($D\lambda^\alpha$ term) vanishes identically making  the corresponding second order metric equations become identical to (\ref{coframe-eqn-Q-palatini}). One thus concludes that quadratic curvature models of type $f(P)$ in both first order and metric theories admit a common set of Einstein metric solutions. However, the number of dimensions is an important parameter,
since for example, in three dimensions any Einstein metric has  constant curvature. Moreover, $*T^\alpha_P$ has vanishing trace only in four dimensions and otherwise its trace is proportional to the density $\mathcal{L}_P.$

In the first order formalism and in the context of ST equivalence, the trace of the coframe equations yield $f-Pf'=0$ and thus the consistency of the equations requires the Legendre transform of  $f(P)$ to vanish and they do have a scalar-tensor equivalent with a dynamical scalar field as in the case of the metric framework.

It is easy to show that the universality property of the quadratic curvature Lagrangian $f(P)*1$ is also shared by $f(Q)*1$ \cite{borowiec0}.
Take, for simplicity of the argument,  the Lagrangian $Q*1$. In the first order formalism and with the assumption $R^\alpha= k\,\theta^\alpha$, the corresponding auxiliary 2-form then becomes $X^{\alpha\beta}_Q=8k^2\,\theta^{\alpha\beta}$ and both the connection and the coframe equations are satisfied identically for a Riemannian connection. Consequently, the universality property of the $f(P)*1$  in fact can be extended to the Lagrangian densities of the form $f(Q,P)*1$.

For the case $f$ is a nonlinear function of $P$, the connection equations allow  an algebraic torsion, cf. Eqn. (\ref{algebraic-torsion}),
as in the case of Einstein-Cartan type  $f(R)$ theory discussed in Sec. III.

Consequently, the connection $\Gamma_{\alpha\beta}$ has exactly the same decomposition as (\ref{f(R)-connections-relation}) with $f'=f_P$ and
the coframe equations assume the form
\be
\Omega_{\mu\nu}(\Gamma)\wdg i^\alpha *X^{\mu\nu}-\tfrac{1}{2}i^\alpha\mathcal{L}_Q=0,
\ee
which can be written in terms of the  geometrical quantities belonging to the metric compatible and torsion-free connection $\omega$ related to $\Gamma$ by (\ref{algebraic-torsion2}).

As in the case of the $R^2$ Lagrangian, the conformal transformation of the Ricci-squared Lagrangian and the corresponding field equations do not lead to a simplified
form  for ST equivalents. Because the expressions involved are relatively more manageable, it is preferable to work at the Lagrangian level to find how Ricci-squared Lagrangian transforms under conformal transformations.

Under the conformal transformation defined by $g\mapsto\tilde{g}=\phi^{-2}g$ or equivalently, $\tilde \theta^\alpha=\phi^{-1}\theta^{\alpha}$ in terms of coframe basis 1-forms in $n$ dimensions, the transformation of the Ricci-squared Lagrangian density can be expressed  in the form
\begin{widetext}
\begin{eqnarray}
\tilde{\mathcal{L}}_P\nonumber
&=&
\tilde{R}^{{\mu}} \wedge \tilde{*} \tilde{R}_{\mu}=\tilde{R}^{{\mu\nu}}\tilde{R}_{{\mu\nu}}\tilde{*} 1 
=
\phi^{(n-4)} \bigg\{R^{\mu\nu} R_{\mu\nu}+2 (2-n) \big[ D_{\nu}(d\ln \phi)_\mu -(d\ln \phi)_\nu(d\ln \phi)_\mu \big] R^{\mu\nu}    \bigg. \nonumber  \\
&+&
\big[(2-n) (d\ln\phi)_\mu (d\ln\phi)^\mu-D_{\mu} (d\ln\phi)^\mu \big] R  +(2-n)^2 D_{\mu}(d\ln \phi)^\nu D^{\mu}(d\ln \phi)_\nu  \nonumber \\
&-&
2 (2-n)^2 D^\nu (d\ln\phi)_\mu (d\ln \phi)^\mu (d \ln \phi)_\nu- (2-n)^2 (n-1) \left[(d\ln\phi)_\alpha (d\ln\phi)^\alpha \right]^2 \nonumber \\
&+&
 (3n-4) \left[D_{\mu} (d\ln\phi)^\mu \right]^2+ 2 (2-n)(3-2n) D^{\mu}(d\ln\phi)_\mu (d\ln\phi)_\nu (d\ln\phi)^\nu
\bigg\} *1,  \label{ctr}
 \end{eqnarray}
\end{widetext}
where the shorthand notation $D_\mu\equiv i_\mu D$ has been used.
Apparently, with a conformal transformation  (\ref{ctr}), the Ricci-squared Lagrangian $\mathcal{L}_P$ cannot be transformed into an  Einstein-Hilbert
Lagrangian with a nonminimally coupled scalar field in a simple manner as in $f(R)$ models \cite{dabrowski}.

\subsection{Modified quadratic curvature  Lagrangians in a cosmological context}

In order to account for the late-time expansion rate of the Universe, the gravitational Lagrangians that modify the usual Einstein-Hilbert Lagrangian
provide  popular alternatives to the general relativistic cosmological models that involve dark energy and dark matter.
Yet another approach is to alter the gravitational sector of the standard cosmological models so that it reconciles with the current cosmological observations. The ST equivalence for modified gravitational Lagrangians introduced above blurs the distinction between the two approaches.
In this regard, it may be worthwhile to isolate the properties and predictions  of some  favorable modifications of the geometrical sector for a given cosmological model which cannot be reduced to simpler models having  scalar degrees of freedom.

Although they are in conflict with solar system tests for gravity, the modified gravitational Lagrangians, for example, of the $f(R)$ form,
\be\label{carroll-lag}
f(R)
=
R-\frac{\mu^{4}}{R},
\ee
are considered in attempts to account for the observed late-time cosmic  acceleration.
Here $\mu$ is a parameter which has the  dimension of mass \cite{carroll1}.
The basic idea behind  the addition of $R^{-1}$ to the Einstein-Hilbert Lagrangian (\ref{carroll-lag})  is to change the field equations in the low curvature  regime. Although it is possible to introduce the ST equivalent of (\ref{carroll-lag}) in the way highlighted in Sec. III which brings (\ref{carroll-lag})
to Einstein gravity with scalar field  field coupled minimally to gravity (and nonminimally coupled to the matter fields if present), slightly more
general modified gravitational Lagrangians  are introduced later \cite{carroll2}.
In a similar manner, in order to investigate the cosmological consequence of vacuum   models by modifying the  gravitational sector only,
inverse powers of quadratic curvature invariants have been adopted.
 In particular, (\ref{carroll-lag})  is generalized to include the additional  quadratic curvature scalars $Q, P, K$  in the form,
\be\label{carroll2}
f(R, Q, P, K)
=
R-\frac{\mu^{4n+2}}{(aQ+bP+cK)^n},
\ee
where $n$ is an integer. This form avoids the flat space solution and has desirable dynamical features  similar  to the Lagrangian in (\ref{carroll-lag}) in connection with the cosmic acceleration \cite{carroll2}. The ST equivalent Lagrangian for (\ref{carroll2}) has been introduced  before in a a different form in \cite{chiba} in connection with its particle content when linearized around a curved background.
It is argued that such  modified gravitational Lagrangians are not conformally equivalent to Einstein gravity plus scalar matter sources \cite{carroll2} so that
such models reflect solely the effects of modification of the gravitational sector. The multi-ST  equivalent for (\ref{carroll2}),  on the other hand, offers  a somewhat simplified fourth order model with a nonminimally coupled single scalar degree of freedom.

 For the sake of simplicity, here   the modified Lagrangian density of  the form
\be\label{qc-modified}
\mathcal{L}_{mod. qc}
=
(aQ+bP+cK)^m*1
\ee
with $m$ is an integer will be considered. For this particular subclass of Lagrangians, the Legendre transform introduced above simplifies considerably to yield an explicit form of the potential term in terms of the scalar field. In this case, for the specific form of the function $f$ in (\ref{qc-modified}), $f_Q, f_P,$ and $f_K$ are all proportional to one another and thus it is possible to introduce a single scalar field coupled nonminimally to quadratic curvature Lagrangians.
Explicitly, the ST equivalent for (\ref{qc-modified}) becomes
\be\label{qc-st-eqiv}
\mathcal{L}_{eq.}
=
\phi\mathcal{L}_{qc}*1-V(\phi)*1,
\ee
where, for convenience, the quadratic curvature Lagrangian $\mathcal{L}_{qc}$ is defined as
\be\label{qc-lag}
\mathcal{L}_{qc}
\equiv
(aQ+bP+cK)*1,
\ee
and the scalar field is obtained by the field redefinition
$
\phi\equiv(aQ+bP+cK)^{m-1}.
$
Consequently, the potential term in (\ref{qc-st-eqiv}) takes the form of a fractional power law as
\be
V(\phi)
=
\frac{(m-1)}{m}\phi^{m/(m-1)}.
\ee
Therefore, the field equations for ST equivalent Lagrangian (\ref{qc-st-eqiv}) can be derived from those of (\ref{qc-lag}) by proceeding in  the same way as in the first example above. The field equations for (\ref{qc-lag}) can be written in the form
\be
*E^\alpha_{qc}
=
D\lambda_{qc}^\alpha+*T^{\alpha}_{qc}=0,
\ee
where both terms are to be calculated by using the auxiliary tensor-valued 2-form
\be
X^{\alpha\beta}_{qc}
=
2a\,R\theta^{\alpha\beta}+b(\theta^{\alpha}\wdg R^{\beta}-\theta^{\beta}\wdg R^{\alpha})+2c\,\Omega^{\alpha\beta}.
\ee
Then the corresponding 2-form for (\ref{qc-st-eqiv}) is $X^{\alpha\beta}_{eq.}=\phi X^{\alpha\beta}_{qc}$. Consequently, the Lagrange multiplier 2-forms
for (\ref{qc-st-eqiv})  and (\ref{qc-lag}) are related by
\be
\lambda_{eq.}^\alpha
=
\phi \lambda_{qc}^\alpha
+
2(d\phi)_\beta *X^{\alpha\beta}_{qc}.
\ee
Eventually, one finds the equation
\begin{align}
\phi *E^{\alpha}_{qc}
+
2D[(d\phi)_\beta*X^{\beta\alpha}_{qc}]
+
2d\phi\wdg i_{\beta} D*X^{\beta\alpha}_{qc}
\nonumber
\\
+
\tfrac{(1-m)}{m}\phi^{m/(m-1)}*\theta^\alpha=0,
\end{align}
which is a simplified and special case of the general expression (\ref{general-modified-eqn}). By tracing the field equations, after some algebra, one finds the equation for the scalar field
\begin{align}
(6a+2b)\,R\,d*d\phi
+
[4(6a+2b)+2c]\,d\phi\wdg *dR
\nonumber\\
+
(6a+2b+c)\phi\, d*dR
+
2(b+c)\,d*[(d\phi)_{\alpha}R^\alpha]
\nonumber
\\
+
\tfrac{4(1-m)}{m}\phi^{m/(m-1)}*1
=0.
\end{align}
Note that it is possible to recover the subcases mentioned above referring \cite{baykal} with particular values of the coupling constants $a,b,c$  for which  $\phi$ is nondynamical.

The subcase corresponding to $m=-1$ in (\ref{qc-modified}) as well as the individual inverse powers of the quadratic curvature invariants supplemented with Einstein-Hilbert action have been studied  in  \cite{carroll2}
in the context of cosmological models. In that work, it is stated that  the generalization of the gravitational Lagrangian of the form (\ref{qc-lag}) to the modified form  (\ref{qc-modified}) is not equivalent to Einstein  gravity plus  matter sources. What we see in this work is that any modification of the form $f(Q, P, K\ldots)$  induces scalar fields coupled nonminimally to individual curvature scalars $Q, P, K,\ldots$   in  ST theory of the  form (\ref{qc-lag})
in the same sense that $f(R)$ models have ST equivalents. Thus,  modifications of gravitational Lagrangian inevitably induce nonminimal scalar matter couplings and in this sense the  ST equivalence blurs the distinction between  adding matter fields, for example of the form  dark energy/matter to Einstein equations and changing the gravitational sector starting from a modified gravitational action.
 In particular, in the case of the gravitational Lagrangian of the form (\ref{carroll2}), the scalar-tensor equivalent turns out to be a simpler
gravitational model with a single scalar field coupled nonminimally to curvature components and having  a suitable potential term for the resulting scalar field.

\subsection{Modified Weyl gravity}
For conformally invariant   Weyl gravity \cite{bach,dereli-tucker-weyl}, the Lagrangian density can be written in terms of the contraction of the Weyl 2-form $C^{\alpha\beta}$ with itself in the following convenient form
\begin{align}\label{weyl-square-eqn-action}
\mathcal{L}_{W}
&=
C_{\alpha\beta}\wdg*C^{\alpha\beta}
\nonumber\\
&=
\Omega_{\alpha\beta}\wdg*\Omega^{\alpha\beta}
-
\tfrac{1}{2}R_\alpha\wdg *R^\alpha
+
\tfrac{1}{6}R^2*1,
\end{align}
where  Weyl 2-form $C^{\alpha\beta}$ is the trace-free part of the curvature 2-form and can be expressed as
\be
C^{\alpha\beta}
=
\Omega^{\alpha\beta}
-
\tfrac{1}{2}(\theta^\alpha\wdg R^\beta-\theta^\beta\wdg R^\alpha)
+
\tfrac{1}{6}R\,\theta^{\alpha\beta}.
\ee
With the help of the result that $\lambda_W^\alpha=2C^\alpha$ where $C^\alpha= DL^\alpha$ is the Cotton 2-form derived from Schouten 1-form $L^\alpha=R^\alpha-\tfrac{1}{6}R\,\theta^\alpha$  \cite{cotton} and using the above general formulas for quadratic curvature Lagrangians, it is possible to show that the field equations that follow from (\ref{weyl-square-eqn-action}) take the form
\be\label{n-dim-weyl-eqns}
-R_\beta\wdg *C^{\alpha\beta}
+
D*C^\alpha
+
\tfrac{1}{2}*T^\alpha_W
=0.
\ee
The last term explicitly has the form
\be\label{energy-momentum-weyl}
*T^\mu_W
\equiv
-(i^{\mu}C^{\alpha\beta})\wdg*C_{\alpha\beta}+C^{\alpha\beta}\wdg i^{\mu}*C_{\alpha\beta},
\ee
and vanishes identically in four dimensions \cite{lovelock}.  In the present notation, this result can simply be obtained by making use of the identity
$*C^{\alpha\beta}=\frac{1}{2}\epsilon^{\alpha\beta}_{\phantom{aa}\mu\nu}C^{\mu\nu}$ satisfied by Weyl 2-form in four dimensions (the same identity is also used to simplify the Lagrange multiplier term as well, see \cite{cotton}) in the expression (\ref{energy-momentum-weyl}). Consequently, the Weyl gravity equations further reduce to
\be\label{simplified-weyl-eqn}
*E^\alpha_W
=
R_\beta\wdg *C^{\beta\alpha}
+
D*C^\alpha
=0.
\ee
Note here that, in the component form the left hand side  of (\ref{simplified-weyl-eqn}) defines the Bach tensor $B_{\alpha\beta}$ \cite{bach}. In the present notation, and thus relative to an orthonormal coframe, it can be defined  by introducing the vector-valued form $B^\alpha=E^\alpha_W$ by means of  the identification $B^{\alpha}=B^{\alpha}_{\fant{a}\beta}\theta^\beta$.

Now consider  the field equations for the modified gravitational Lagrangian based on the Weyl scalar $C$ which can be defined as $C*1\equiv C_{\alpha\beta}\wdg*C^{\alpha\beta}$. In the spirit of the $f(R)$  model, consider the modified Lagrangian  of the form
\be
\mathcal{L}_{mod.W}
=
f(C)*1.
\ee
By making use of the field redefinitions  $\frac{df}{dC}\equiv \phi$, one has $\Pi^{\alpha\beta}_{mod.W}=2\phi*C^{\alpha\beta}$ and therefore,
\be
\lambda_W^\alpha
=
2\phi*C^\alpha +4(d\phi)_\beta*C^{\alpha\beta}.
\ee
Consequently, the field equations take the form
\be\label{modified-weyl-eqn}
\phi *B^\alpha
+
2d\phi\wdg i_\beta D*C^{\alpha\beta}
+
2D[(d\phi)_\beta*C^{\alpha\beta}]
-
V(\phi)*\theta^\alpha
=0,
\ee
where the potential for the scalar is, as before, given by the Legendre transform $V(\phi)=f-C \phi$
with $\phi\equiv\frac{df}{dC}$ and note also that $(d\phi)_\beta$ refers to the components of the 1-form $d\phi$ relative to an orthonormal coframe.
The field equations for Weyl theory (\ref{simplified-weyl-eqn}) have  vanishing trace, $B^\alpha_{\fant{a}\alpha}=0$ as a consequence of
trace-free property of Bach form which in turn follows from the common properties of Weyl 2-form and Cotton 2-form, that is $i_\alpha C^{\alpha\beta}=0$ and $i_\alpha C^{\alpha}=0$ respectively.
This property is also maintained in the modified equations and thus leads to the fact that the trace of the vacuum field equations
(\ref{modified-weyl-eqn}) implies  the constraint
\be\label{weyl-constraint}
f(C)-Cf'(C)=0,
\ee
i.e., the potential for the scalar vanishes for consistency of the vacuum field equations. Thus, in this case the scalar field turns out to be
 nondynamical. Consequently, for the modified Weyl Lagrangian there is no advantage in field redefinitions via a Legendre transform.
 In terms of $f'=\frac{df}{dC}$, the vacuum field equations (\ref{modified-weyl-eqn}) can be rewritten as
\be\label{modified-weyl-eqn-form2}
f'*B^{\alpha}
+
2df'\wdg i_\beta D*C^{\alpha\beta}
+
2D[(df')_\beta*C^{\alpha\beta}]
=0,
\ee
subject to the condition (\ref{weyl-constraint}).
With regard to the practical use of the resulting field equations   to the problems of interest in black hole physics for a given $f(C)$, (\ref{modified-weyl-eqn-form2}) may, for example, allow one to adopt an alternative and direct approach to the analysis given in \cite{frolov} provided that the above analysis is extended to higher dimensions.

\section{A Modified Sixth Order Lagrangian}

The above procedure of introducing ST equivalents can also be extended to even higher order theories which was studied previously
in \cite{mazumdar}.
For example, gravitational  Lagrangians involving cubic Ricci terms lead to sixth order  metric equations.
This section deals with another type of sixth-order
gravitational Lagrangian based on the square of the gradients  of the scalar curvature. In terms of differential forms such a Lagrangian form  can be written as
\be\label{6th-order-lag}
\mathcal{L}_s
=
dR\wdg *dR.
\ee
For (\ref{6th-order-lag}),  one has $\Pi_{s}^{\alpha\beta}=D*[(\Delta R) \theta^{\alpha\beta}]$, and using the general formulas of Sec. 2, one can easily  show that it leads to the sixth order vacuum field equations
\be\label{6th-order}
*E^\alpha_s
=
-
4(\Delta R)*R^\alpha
+
2D*(\Delta dR\wdg \theta^\alpha)
-
*T^\alpha_s=0.
\ee
 Equation (\ref{6th-order}) involves the Laplace-Beltrami operator $\Delta$ defined above and the Lagrange multiplier term,  the second term in (\ref{6th-order}), contains sixth order partial derivatives of the metric components relative to a coordinate coframe. 3-form $*T^\alpha_s$ is defined as
\be\label{en-mom-T_s}
*T^\alpha_s
\equiv
(i^\alpha dR)*dR+dR\wdg i^\alpha*dR,
\ee
similar to the energy-momentum 3-form for the scalar field (\ref{scalar-en-mom}).  Unlike the above quadratic curvature cases, 3-form $*T^\alpha_s$ is not calculated by a corresponding auxiliary 2-form $X^{\alpha\beta}$, but it results from the commutation of the variational derivative with $*$ for (\ref{6th-order-lag}). However, eventually, the field equations written as (\ref{6th-order}) take a form similar to the those studied above in Sec. III. In  fact, Eq. (\ref{6th-order}) can be brought into  form more akin to those of a Brans-Dicke type ST   theory as in  the $f(R)$ case.

 By comparing the field equation (\ref{6th-order}) with (\ref{ohanlon-eqns}) it is possible to deduce that
(\ref{6th-order}) can be written in the form analogous to (\ref{ohanlon-eqns}) by the simple field redefinition $\Delta R\equiv\phi.$
Note that such a mathematical connection is not obvious at the level of Lagrangian.
In doing so, (\ref{6th-order}) then takes the form
\be\label{st1-6-th-order}
-4\phi*R^\alpha-*T^\alpha_s+2D*(d\phi\wdg \theta^\alpha)=0.
\ee

By taking the trace of Eq. (\ref{6th-order}), one finds
\be\label{trace-6th-order}
E^{\alpha}_{s\alpha}*1
=
3\Delta\Delta R*1+R\Delta R*1-dR\wdg*dR=0,
\ee
or equivalently, in terms of the scalar field it can be rewritten in the form
\be
3d*d\phi +R\phi*1-dR\wdg*dR=0.
\ee
The frame in  (\ref{6th-order}) or (\ref{st1-6-th-order}) can be regarded as Jordan frame equations. One can try to put these into Einstein frame equations by considering a conformal transformation where equations coupled minimally to a scalar field with additional curvature-source terms.

The model based on $f(R)*1+a dR\wdg *dR$ has been studied before in the context of cosmological perturbations \cite{hwang,mazumdar}. By making use of the field equations (\ref{BD-type-eqn}) and (\ref{6th-order}), it is possible to write the metric equations of such a Lagrangian in a form similar to (\ref{BD-type-eqn}), explicitly  in the form
\be\label{staro-hj-example}
-2F*G^\alpha+2D*(dF\wdg\theta^\alpha)+ (f-RF)*\theta^\alpha-a*T^\alpha_s=0,
\ee
where the definition $F\equiv f_R-2a\Delta R$ has been introduced for convenience.
In particular, for $f(R)=R+bQ$, it was shown that, by suitable field redefinitions after a conformal transformation analogous to the case for $f(R)$ models, it is possible to rewrite (\ref{staro-hj-example}) as Einstein gravity with two interacting scalar fields \cite{starobinsky-hj-schmidt}.

It is now a convenient point to introduce  ST equivalents for the generalization of the Lagrangian (\ref{6th-order}) in $f(R)$ fashion.
One proceeds  in the same way as in the examples before and first define the scalar $S$ via $S*1\equiv dR\wdg*dR $. Then, the formulas of previous sections  can be used to derive the explicit form of the field equations based on the modified Lagrangian
\be
\mathcal{L}_{mod.s}=f(S)*1,
\ee
by making use  of the field equations (\ref{6th-order}). Explicitly, by introducing $\frac{df}{dS}\equiv f'\equiv \phi$, they can be written in the form
(\ref{general-modified-eqn}) as
\begin{align}
\phi*E^\alpha_s
+
V(\phi)*\theta^\alpha
+
2\phi D*(\Delta dR\wdg \theta^\alpha)
\nonumber
\\
+
2d\phi\wdg *(\Delta dR\wdg \theta^\alpha)
+
2\Delta dR\wdg *(d\phi\wdg \theta^\alpha)
=0,
\end{align}
where the potential term is the Legendre transform  $V(\phi)\equiv f-Sf'$. The trace of the modified field equations then yields
a second order  dynamical field equation for the scalar field coupled to the derivatives of the scalar curvature $R$ as
 \be
 6(\Delta\phi)\Delta R*1
 -
 12\Delta dR\wdg *d\phi
 +
 \phi E^{\alpha}_{s\alpha}*1
  +
 4V(\phi)*1
 =0.
 \ee
Finally, it is a worthwhile to note that
this example also illustrates a subtlety  related to surface terms in the derivation of the field equations for modified gravity models. Recall that,
using the properties of exterior derivative, coderivative and the Laplace-Beltrami operators, one has
\be
d(R*dR)=-R\Delta R*1+dR\wdg *dR.
\ee
Thus, (\ref{6th-order-lag}) is equivalent to the Lagrangian density $R\Delta R*1$ and consequently they
lead to the same field equations. On the other hand, it is easy to show that the field equations that follow from $f(R\Delta R)*1$ are not equivalent to those that follow from $f(*(dR\wdg*dR))*1$.

\section{Concluding comments}

An apparent advantage of the  use of the first order formalism is that it allows one to treat various modified gravitational Lagrangians ranging  Ricci-squared Lagrangian in Palatini formalism and $f(R)$ models with nonmetricity and torsion to modified curvature-squared Lagrangians in a unified framework.
From a technical point of view, it is also possible to relate the first order formalism to the C-theories introduced   recently in \cite{amendola}.
In addition, the use of exterior algebra of tensor-valued forms  renders the tensorial manipulations easier compared to the methods involving corresponding tensor components.

A general modified gravitational Lagrangian of the form $f(R, P, Q, K, S)*1$  can be reduced to simpler Lagrangians involving $R, P, Q, K, S,\ldots$
with a number of nonminimally coupled scalar fields in the sense that $f(R)*1$ is related to the Einstein-Hilbert Lagrangian $R*1$ with nonminimally coupled scalar fields. The dynamical degree of freedom for the scalar fields, as well as the number of resulting independent scalar fields depend on the explicit form of the function $f$ as discussed with the explicit examples.

The multi-scalar-tensor  equivalence discussed  above for the modified gravitational actions  can be regarded as an extension of the equivalence between the Brans-Dicke-type ST theory and $f(R)$ theory to more complicated gravitational Lagrangians.  In this  regard, the procedure  generates  Brans-Dicke type scalar fields each of which are nonminimally coupled to curvature components.  On the other hand $f(R)$ models are simple enough to accommodate a general connection with algebraic (i.e. nonpropagating) torsion and nonmetricity. These non-Riemannian models can also be cast into a form expressed in terms of pseudo-Riemannian quantities in addition to new gravitational interactions induced by non-Riemannian parts \cite{dereli-tucker}. These features of $f(R)$ models, in general, do not carry over to modified gravitational Lagrangians based on more complicated curvature scalars.

The important issues to be addressed in constructing  multi-ST equivalents are the number of resulting scalar fields  and the determination of the  dynamical degrees of freedom by the resulting scalar fields.
The former  problem has been addressed  in \cite{starobinski-shapiro} by using constraint analysis and the theory of primary constraints
 related to the Legendre transform of constrained systems. The latter issue can be addressed by discussing the ST equivalents at the level of field equations as we have discussed above. In the case where the scalar field is nondynamical ST equivalence becomes a superficial field  redefinition.
Moreover, save  for some possible special cases, the  scalar field satisfies a field equation of its own with  certain curvature interaction terms depending  on the particular curvature scalars under consideration.

With the experience gained through the ST equivalents for the gravitational models explicitly discussed above, the scope of the ST equivalence  can be expanded to cover topological terms in gravitational models. In fact, the application of the idea of ST equivalence to topological terms provides a mechanism
to incorporate such terms into gravitational Lagrangians \cite{baykal-delice}.
Consider, for example, the Chern-Simons modified gravity where the Einstein-Hilbert Lagrangian is supplemented with a  term
$\Omega^{\alpha}_{\fant{a}\beta}\wdg\Omega^{\beta}_{\fant{a}\alpha}$ multiplied by a so-called cosmic scalar field.
It is well known that, without a multiplicative  scalar field, such a term does not contribute to the field equations, since it is the exterior derivative of the term
$
d\omega^{\alpha}_{\fant{a}\beta}\wdg\omega^{\beta}_{\fant{a}\alpha}
+
\tfrac{2}{3}\omega^{\alpha}_{\fant{a}\beta}\wdg\omega^{\beta}_{\fant{a}\mu}\wdg\omega^{\mu}_{\fant{a}\alpha}.
$
As for all the examples studied above, one starts with the scalar $T$ which is defined by $T*1\equiv \Omega^{\alpha}_{\fant{a}\beta}\wdg\Omega^{\beta}_{\fant{a}\alpha}$, the Lagrangian density of the form $f(T)*1$     with arbitrary function $f$ contributes to the field equations without the need to introduce a scalar field.
In the same manner, in four dimensions, it is possible to incorporate quadratic Gauss-Bonnet term
$G*1\equiv \Omega_{\alpha\beta}\wdg \Omega_{\mu\nu}\wdg *\theta^{\alpha\beta\mu\nu} $
 into a gravitational Lagrangian by considering the term of the form $f(G)*1$ see, for example, \cite{baykal-delice}. However, in neither of the  cases the scalar field  is dynamical.

\section*{Acknowledgment}
It is a pleasure to dedicate the paper to Professor Metin Ar\i k on the occasion of his 65th birthday.

\section*{Appendix}

In order to facilitate comparison with the existing  literature, often involving expressions and computations relative to a coordinate
coframe,  an illustration of how field equations relative to a coordinate coframe can be derived from the corresponding equations relative to
an orthonormal  coframe will be presented in some detail.

Any of the field equations $*E_\alpha=0$ derived above in terms of the  vector-valued form  $E_\alpha=E_{\alpha\beta}\theta^{\beta}$
is related  to the  coordinate expression $E_{ab}=0$ by $E_{\alpha\beta}$  with  $E_{\alpha\beta}=e^{a}_{\alpha}e^{b}_{\beta}E_{ab}$.
(Here the Latin indices refer the to components of tensors relative to a coordinate coframe). Likewise, the basis coframe 1-forms  are related by
$\theta^{\alpha}=e^{\alpha}_adx^a$ whereas the basis frame fields are related by $e_{\alpha}=e^{a}_{\alpha}\partial_a$ with
$e^{\alpha}_{a}e^{a}_{\beta}=\delta^{\alpha}_{\beta}$ and $e^{\alpha}_{a}e^{b}_{\alpha}=\delta^{a}_{b}$.

After these preliminary definitions,
take, for example  the metric vacuum field equations $*E^\alpha=0$ for the $f(R)$ model,   (\ref{BD-type-eqn}) derived above relative to an orthonormal coframe.
These    can conveniently be rewritten in the form
\be\label{metric-f(r)-on-version}
-
*\left(f'R^{\alpha}-\frac{1}{2}f\theta^\alpha\right)
+
D *(df'\wdg \theta^{\alpha})
=0.
\ee
The derivation of the corresponding coordinate expressions then involves converting covariant exterior derivatives $D$ in the second term into covariant derivatives
$\nabla_{e_a}\equiv\nabla_{a}$. This requires careful use of the properties of the covariant exterior derivative. In that,  note that although $\nabla_a$ commutes with the Hodge dual operator $*$, $D$ does not commute with it.
In particular, this term can be put into a more common form by using the operator identity,
\be\label{last-id}
Di_a+i_aD=\nabla_a,
\ee
acting on an arbitrary  $p$-form.
For a derivation of this crucial identity as well as other  properties of the covariant exterior derivative, the reader is referred to e.g. \cite{tucker}.

In order to make use of the identity  (\ref{last-id}),
 it is convenient to write the second term  in the form
$D *(df'\wdg \theta^{\alpha})=D i^\alpha*df'$. Here,  another identity $i_{e^\alpha}*\sigma=*(\sigma\wdg \theta^\alpha)$ is used. This identity involving Hodge dual and the contraction operator holds for an arbitrary $p$-form $\sigma$ and the basis frame fields with a superscript are defined by $e^\alpha=\eta^{\alpha\beta}e_\beta$.
Consequently, one ends up with
\be
D *(df'\wdg \theta_{a})
=
\nabla_a *df'
-
i_a(d*df'),
\ee
where the property that the covariant exterior derivative acting on an arbitrary form   reduces to the exterior derivative has been used.
Finally, by  noting that covariant derivative of a Riemannian connection commutes with the Hodge dual operator $*$,
which explicitly be stated as
\be
\nabla_\alpha*=*\nabla_\alpha,
\ee
and that $\Delta f'\equiv *d*df'$, the second term simplifies to
\be
D *(df'\wdg \theta_{\alpha})
=
*[\nabla_\alpha df'
-
(\Delta f')\theta_\alpha].
\ee
On the other hand, it follows from the definition of covariant derivative that  $\nabla_{e_\alpha}=e^{a}_{\alpha}\nabla_{\partial_a}$. Consequently,
it is possible to read off the coordinate expressions from the following expression
\be
D *(df'\wdg \theta_{\alpha})
=
e^{a}_{\alpha}*[\nabla_a df'
-
(\Delta f')g_{ab}dx^b],
\ee
since the expression in front of the Hodge dual now involves tensorial quantities relative to a coordinate coframe.

Eventually, this result allows one to extract  $E_{ab}=0$  from the expression $E_{ab}*dx^b=0$ which simply  follows from $*E_{a}= e^{\alpha}_{a}*E_{\alpha}$.
Explicitly, by dropping the irrelevant Hodge dual,  the resulting equations read
\be\label{coord-eqn1}
E_{ab}
=
f'R_{ab}-\frac{1}{2}g_{ab}f-\nabla_{a}\nabla_b f'+g_{ab}\nabla^c\nabla_c f'=0,
\ee
 relative to a coordinate frame. Evidently,
these equations can directly be derived
from the invariant Lagrangian density
\be
f(R)*1
=
f(R)\sqrt{|g|}dx^{0}\wdg dx^{1}\wdg\cdots \wdg dx^{n},
\ee
where $*1=\theta^{012\cdots n}=|e|dx^{01\cdots n}=\sqrt{|g|}dx^{01\cdots n}$ is the oriented volume element relative to a coordinate basis.
These results also allow one to write
the coordinate expression for the field equations of the ST equivalent of (\ref{coord-eqn1}) immediately
in the form
\be
\phi\, G_{ab}-\nabla_{a}\nabla_b \phi +g_{ab}\nabla^c\nabla_c \phi-\frac{1}{2}g_{ab}V(\phi)=0.
\ee
These metric equations follow from the following invariant Lagrangian density
\be
L_{ST}
=
\left(\phi \, g^{ab}R_{ab}-V(\phi)\right)\sqrt{|g|}dx^0\wdg dx^1\cdots \wdg dx^n,
\ee
where $R_{ab}$ are the components of the Ricci tensor relative to a coordinate coframe.

The field equations for the $f(R)$ model with nonmetricity and  torsion also have similar terms and the corresponding coordinate expressions can be written
in a similar way. Most of the identities  used in the above  calculations in deriving coordinate coframe expressions for a given  modified gravitational Lagrangian are also useful in deriving coordinate expressions  involving the Lagrange multiplier terms as well.

Likewise, in Section IV, the field equations for the sixth order gravity based on  the Lagrangian density
\be
dR\wdg *dR
=
g^{ab}\partial_a R \partial_b R\sqrt{|g|}dx^{1}\wdg\cdots \wdg dx^{n},
\ee
is considered where  the coordinate expression on the right-hand side follows from
the exterior derivative $dR=\partial_a R dx^a$ and consequently
the contraction
of the basis coframe 1-forms as
\be
dx^a\wdg *dx^b
=
g^{ab}\sqrt{|g|}dx^{1}\wdg\cdots \wdg dx^{n}.
\ee
It is possible to derive the coordinate expression for the sixth  order model by making use of the expression $E_{ab}$ for the $f(R)$  model
since  the corresponding field equations are cast in a form that is formally similar to those of the $f(R)$ model.
It is sufficient to derive the coordinate expression for the vector-valued form $*T^\alpha_s$ defined in (\ref{en-mom-T_s}), which can be rewritten in the form as
\begin{align}
*T^\alpha_s
&=
(i^\alpha dR)*dR+i^\alpha(dR\wdg *dR)
\nonumber\\
&=
e^{\alpha}_a\left[2(i^a dR)*dR-i^a(dR\wdg *dR)\right],
\end{align}
and therefore, by recalling that $i_a d R=\partial_a R$  relative to a coordinate coframe, one can read off
the coordinate expression  for the energy momentum 1-form as
\be
T_a^s
=
2\partial_aRdR-g_{ab}g^{cd}\partial_c R\partial_d R dx^b,
\ee
 and in turn, using this result, the coordinate expression for $T_{ab}^s$
can be obtained as
\be
T^s_{ab}
=
2\partial_a R\partial_b R
-
g_{ab}g^{cd}\partial_c R\partial_d R.
\ee
This result then allows one to write the explicit form  of $E_{ab}=0$ for the sixth order model as
\begin{align}
&FR_{ab}
-
\frac{1}{2}g_{ab}F-\nabla_{a}\nabla_b F+g_{ab}\nabla^c\nabla_c F
\nonumber\\
&-
\partial_a R\partial_b R
+
\frac{1}{2}g_{ab}g^{cd}\partial_c R\partial_d R=0
\end{align}
with $F=\Delta R=g^{ab}\nabla_a\nabla_b R$ replacing  the function $f'$ for the $f(R)$ model as  a consequence of the identity $d\Delta=\Delta d$.
For the coordinate expression for the ST equivalents related to the sixth order model, the reader is referred to the original work \cite{starobinsky-hj-schmidt}.

The modified-QC gravity field equations are based on the quadratic curvature invariants of the form
\be
L=f\left(R^2,R_{ab}R^{ab},R_{abcd}R^{abcd}\right)\sqrt{|g|}dx^0\wdg \cdots \wdg dx^n,
\ee
and under certain assumptions about the rank of relevant Legendre  transformation, the ST equivalents for this Lagrangian have  the general form
\begin{align}
L_{ST}=\left[\phi_1 R^2+\phi_2 R_{ab}R^{ab}+\phi_3 R_{abcd}R^{abcd}   \right. \nonumber \\
 \left. -V(\phi_1,\phi_2,\phi_3) \right] \sqrt{|g|}dx^0\wdg \cdots \wdg dx^n,
\end{align}
where the independent scalar fields $\phi_1,\phi_2$  and $\phi_3$ are defined in (\ref{multiscalars}) whereas the potential term $V(\phi_1,\phi_2,\phi_3)$ is the Legendre transform of $f(R^2,R_{ab}R^{ab},R_{abcd}R^{abcd})$ given in (\ref{multi-scalar-pot}).
In particular, for the simple modified QC Lagrangians having the particular form
\be
L=\left(aR^2+bR_{ab}R^{ab}+cR_{abcd}R^{abcd}\right)^m\sqrt{|g|}dx^0\wdg \cdots \wdg dx^n,
\ee
there is an ST equivalent with potential for the scalar of the form of a power law
\begin{align}
L_{ST}=&\Big[\phi \left(aR^2+bR_{ab}R^{ab}+cR_{abcd}R^{abcd}\right)
\nonumber\\
&-\left(\frac{1-m}{m}\right)\phi^{m/m-1}
\Big]\sqrt{|g|} dx^0\wdg \cdots \wdg dx^n.
\end{align}

In general QC field equations involve the terms of the form $D*DL^a$, where 1-form $L^a$ is typically linear combination of Ricci 1-form and
the 1-form $Rdx^a$. Take, for example,
the Weyl-squared action in four dimensions. It follows from the invariant Lagrangian density
\begin{align}
C^{\alpha\beta}\wdg*C_{\alpha\beta}
&\equiv
C^{ab}\wdg*C_{ab}
\nonumber\\
&=
\frac{1}{2}C^{ab}_{\fant{ab}cd}C_{ab}^{\fant{ab}cd}
\sqrt{|g|}dx^0\wdg \cdots \wdg dx^3.
\end{align}
The corresponding field equations can be expressed as the vanishing of the Bach tensor $E^W_{ab}\equiv-B_{ab}=0$. An  explicit expression
for the coordinate components of Bach tensor  turns out to be
\be
B_{ab}
=
\nabla^c C_{acb}+R^{cd}C_{acbd},
\ee
where $C_{acb}$ are the components of the Cotton 2-form  \cite{cotton} defined by $C_a=\frac{1}{2}C_{abc} dx^b\wdg dx^c$ whereas $C^{a}_{\fant{a}bcd}$
are components  of Weyl tensor $C^{a}_{\fant{a}b}=\frac{1}{2}C^{a}_{\fant{a}bcd}dx^b\wdg dx^c$.
Moreover, for the divergence of the Cotton 2-form, using the identity (\ref{last-id}) one finds
\begin{align}
D*C^a
&=
\frac{1}{2}DC^{a}_{\fant{a}bc}\wdg *(dx^{b}\wdg dx^{c})
\nonumber\\
&=
\frac{1}{2}\nabla_dC^{a}_{\fant{a}bc} dx^d\wdg *(dx^{b}\wdg dx^{c}).
\end{align}
On the other hand, the contraction on the right-hand side leads to  the following Hodge dual of basis 1-forms
\be
dx^d\wdg *(dx^{b}\wdg dx^{c})
=
-g^{bd}*dx^c+g^{cd}*dx^b.
\ee
Using this result, it is easy to find
\be
D*C_a
=
-\nabla^bC_{abc} *dx^c.
\ee
The Cotton 2-form itself can be expressed in terms of the covariant derivatives of curvature components.
It is derived from Schouten tensor $C^a=DL^a$ where the Schouten tensor $L_{ab}$ can be defined in terms of  $L_a\equiv L_{ab}dx^b$ as  $L_{ab}=R_{ab}-\frac{1}{6}g_{ab}R$  (the definitions in the literature vary up to an overall  constant).
By taking the antisymmetrization induced by the covariant exterior derivative into the account, the coordinate components of
the Cotton 2-form for example take the form
\be
C_{abc}
=
\nabla_{a}R_{bc}-\nabla_{b}R_{ac}-\frac{1}{6}(\nabla_{a}Rg_{bc}-\nabla_{b}Rg_{ac}).
\ee

Evidently, these formulas also help to relate orthonormal components of QC gravity to the corresponding  ones relative to  a coordinate basis.
For a more detailed  account for coordinate components of  a given expression relative to an orthonormal coframe, the reader is referred
to, for example, \cite{cotton}.

\end{document}